\documentclass[final]{ias2}

\usepackage{graphicx}
\usepackage{multirow}
\usepackage{array}

\usepackage{hyperref}

\newcommand{\bea}{\begin{eqnarray}}
\newcommand{\eea}{\end{eqnarray}}
\newcommand{\bes}{\begin{subequations}}
\newcommand{\ees}{\end{subequations}}
\newcommand{\ds}{\displaystyle}

\begin{document}

\markboth{Dynamics of solitons in multicomponent LSRI system}{T Kanna, K Sakkaravarthi, M Vijayajayanthi and M Lakshmanan}

\title{Dynamics of solitons in multicomponent long wave-short wave resonance interaction system}

\author[bhc]{T Kanna}\email{Corresponding author (T. Kanna): kanna\_phy@bhc.edu.in}
\author[bhc]{K Sakkaravarthi}\email{ksakkaravarthi@gmail.com}
\author[anna]{M Vijayajayanthi}\email{vijayajayanthi.cnld@gmail.com}
\author[bdu]{M Lakshmanan}\email{lakshman@cnld.bdu.ac.in}
\address[bhc]{Post Graduate and Research Department of Physics, Bishop Heber College, Tiruchirappalli--620 017, Tamil Nadu, India}
\address[anna]{Department of Physics, Anna University, Chennai--600 025, Tamil Nadu, India}
\address[bdu]{Centre for Nonlinear Dynamics, School of Physics, Bharathidasan University, Tiruchirapalli--620 024, Tamil Nadu, India}

\begin{abstract}
In this paper, we study the formation of solitons, their propagation and collision behaviour in an integrable multicomponent (2+1)-dimensional long wave-short wave resonance interaction ($M$-LSRI) system. First, we briefly revisit our earlier results on the dynamics of bright solitons and demonstrate the fascinating energy exchange collision of bright solitons appearing in the short-wave components of the $M$-LSRI system. Then, we explicitly construct the exact one- and two- multicomponent dark soliton solutions of the $M$-LSRI system by using the Hirota's direct method and explore its propagation dynamics. Also, we study the features of dark soliton collisions.
\end{abstract}



\maketitle

\section{Introduction}
Nonlinear waves appearing in multicomponent nonlinear evolution equations governing the dynamics of various interesting physical systems display intriguing propagation and collision properties. The nonlinear waves, mainly solitons, which arise as the solutions of integrable nonlinear equations show interesting collision features due to their remarkable stability property and find innumerable applications in different areas of science and technology \cite{Whitham-book}. Particularly, higher dimensional multicomponent systems admit various localized structures like solitons, vortex solitons, dromions, and so on. These  multicomponent higher dimensional solitons (HDSs) have attracted our interest to pursue a systematic study on their propagation and intriguing collision dynamics which will be of physical significance in different contexts of nonlinear science. In order to unearth the features of HDS, here we consider the following set of integrable nonlinear evolution equations describing the resonance interaction of multiple short waves (SWs) of high-frequency with a long wave (LW) of low-frequency, which is referred as ($2+1$)-dimensional multicomponent long-wave--short-wave resonance interaction ($M$-LSRI) system,
\bes\bea
&&i (S_{t}^{(\ell)}+ S_{y}^{(\ell)})- S_{xx}^{(\ell)}+ L S^{(\ell)}=0, \qquad \ell=1,2,3,...,M, \\
&&L_{t}=2\sum_{\ell=1}^M |S^{(\ell)}|^2_{x},
\eea\label{model}\ees
where $S^{(\ell)}$ represents the $\ell$-th SW, $L$ indicates the LW and the subscripts represent the partial derivatives with respect to the evolutional coordinate $t$ and the spatial coordinates ($x$ and $y$). In the above (2+1)D $M$-LSRI system, `2' stands for the two spatial dimensions ($x$ and $y$), `1' stands for the evolutional coordinate `$t$', and `$M$' represents the number of SW components of the system.

The resonance interaction of long-wave and short waves takes place when there occurs an exact (approximate) balance between the phase velocity of a LW ($v_p$) and the group velocity of multiple SWs ($v_g$), i.e., $v_p \simeq v_g$ \cite{Zakh1972,Benny,Kawahara1975a,Oikawa1976ptp}. Such LSRI phenomenon in different types of one- and two-dimensional nonlinear systems has been analyzed extensively in the literature (for a detailed information see Refs. \cite{Onorato-prl,Shukla-prl,Funakoshi1989,Ohta2007jpa,Radha2009jpa,Kanna2009jpa,Sakkara2013epjst,Kanna2013pre,Sakkara2014} and references therein). The above mentioned (2+1)D $M$-LSRI system (\ref{model}) is one such model which supports several interesting dynamical features. In the context of nonlinear optics, system (\ref{model}) can be derived from a set of two-dimensional multiple coupled nonlinear Schr\"odinger type equations, when long-wave--short-wave resonance takes place \cite{Onorato-prl,Shukla-prl}.

To highlight the historical perspectives of the considered system, we wish to point out that the simplest form of (\ref{model})- that is, the one component ($M=1$) two-dimensional LSRI system- has been obtained by using a perturbation method in a two-layer fluid model and soliton solutions were constructed by applying the Hirota method \cite{Funakoshi1989}. Latter, in Ref. \cite{Ohta2007jpa}, special bright multi-soliton solutions in the Wronskian form were obtained for the two-component ($M=2$) LSRI system and the Painlev\'e integrability analysis of that two-component LSRI equation was carried out in Ref. \cite{Radha2009jpa} with special dromion solutions. The more general bright multi-soliton solution of the (2+1)D $M$-LSRI system (\ref{model}) was obtained by the present authors in \cite{Kanna2009jpa} and fascinating energy sharing (shape changing) collision of bright solitons have been explored. Also, the propagation and collision dynamics of bright multi-soliton bound states and mixed (bright-dark) solitons of system (\ref{model}) have been discussed in Refs. \cite{Sakkara2013epjst} and \cite{Kanna2012arxiv}, respectively. Recently, new integrable generalizations of $M$-LSRI system (\ref{model}) in (1+1)D, referred as $M$-Yajima-Oikawa system, and in (2+1)D have been reported in Refs. \cite{Kanna2013pre} and \cite{Sakkara2014}, respectively.

The objective of this paper is to showcase the dynamics of bright and dark solitons of the $M$-LSRI system. We obtain the bilinear equations of $M$-LSRI system (\ref{model}) by using the Hirota's direct method in Section \ref{secbilin}. In section \ref{secbright}, we revisit our earlier studies on the dynamics of bright multi-soliton of system (\ref{model}). Then, we construct the one- and two-dark soliton solution of $M$-LSRI system (\ref{model}) and explore its collision dynamics in section \ref{secdark}. We summarize our main results in the final section.

\section{Bilinear Equations of $M$-LSRI system (1)}\label{secbilin}
Hirota's  bilinearization method \cite{Hirota-book} is one of the efficient analytical tools to construct soliton solutions of integrable nonlinear evolution equations due to its algebraic nature. In this section, to obtain the soliton solutions of the $M$-LSRI system (\ref{model}) by applying the Hirota's method, we transform the nonlinear equations (\ref{model}) into a set of bilinear equations using the following transformation
\bes\bea
S^{(\ell)}&=&\frac{g^{(\ell)}}{f},\quad \ell=1,2,...M,\\
L&=&-2\frac{\partial^2}{\partial x^2}(\ln{f}),
\eea \label{trans}\ees
where $g^{(\ell)}$ and $f$ are arbitrary complex and real functions of $x$, $y$ and $t$, respectively.
Then we can write Eqs. (\ref{model}) as a set of bilinear equations:
\bes\bea
&&\left(i(D_t+D_y)-D_x^2\right)g^{(\ell)}\cdot~ f=0,\quad \quad \ell=1,2,3,...,M,\\
&&(D_xD_t-2\lambda) f \cdot~ f=-2\sum_{\ell=1}^{M} \left|g^{(\ell)}\right|^2.
\eea\label{beq}\ees
In the above equations (\ref{beq}), $\lambda$ is an unknown constant to be determined, $D_x$, $D_y$ and $D_t$ are the standard Hirota's $D$-operators \cite{Hirota-book}. For $\lambda=0$, Eqs. (\ref{beq}) admit bright soliton solutions with zero background, while for the general case ($\lambda \neq 0$) Eqs. (\ref{beq}) can exhibit bright-dark and dark-dark soliton solutions. In this paper, we briefly revisit some interesting results of our earlier study on the propagation and collision dynamics of bright multi-solitons \cite{Kanna2009jpa}. Then we construct the dark soliton solutions of $M$-LSRI system (\ref{model}) and investigate their dynamics in detail.

\section{Bright multi-soliton solution and collision dynamics - An overview}\label{secbright}
We have obtained the explicit form of more general bright $n$-soliton solution, for arbitrary $n$, by applying the Hirota's method (see Ref. \cite{Kanna2009jpa}). For this purpose, the power series expansion of variables $g^{(\ell)}$ and $f$ are expressed as
\bea g^{(\ell)}&=&\sum_{j=1}^{n}\chi^{2j-1} g_{2j-1}^{(\ell)}, \qquad \ell=1,2,...M,\nonumber\\ f&=&1+\sum_{j=1}^{n}\chi^{2j} f_{2j}.\nonumber\eea On substituting this $g^{(\ell)}$ and $f$ in the bilinear equations (\ref{beq}) and solving the resulting equations arising at different powers of $\chi$, we get the exact expression for $g^{(\ell)}$ and $f$ in the form of Gram determinants as
\bes\bea
g^{(\ell)}=
\left|
\begin{array}{ccc}
A & I & \phi\\
-I & B & {\bf 0}^T\\
{\bf 0} & a_{\ell} & 0
\end{array}
\right|, \quad \quad f= \left|
\begin{array}{cc}
A & I\\
-I & B
\end{array}
\right|. \label{formgf}
\eea
Then from Eqs. (\ref{trans}) and (\ref{formgf}), the bright $n$-soliton solution can be written as
\bea
S^{(\ell)}&=&\frac{g^{(\ell)}}{f},\quad \ell=1,2,...M,\\
\hspace{-4.0cm}\mbox{and} \qquad \qquad
L&=&-2\frac{\partial^2}{\partial x^2}(\ln{f}).
\eea
In Eq. (\ref{nsol}a), $I$ and $\bf{0}$ represent identity matrix and null matrix of dimensions ($n \times n$) and ($1 \times n$), respectively, $A$ and $B$ are square matrices of dimension ($n \times n$) with elements
\bea
\hspace{-1.6cm} A_{ij}= \frac{e^{\eta_{i}+\eta_{j}^*}}{k_{i}+k_{j}^*},~
B_{ij}=\kappa_{ji}=\frac{-\psi_i^{\dagger} \psi_j}{(\omega_i^*+\omega_j)} \equiv \frac{-\sum_{\ell=1}^M \alpha_j^{(\ell)} \alpha_i^{(\ell)*}}{(\omega_i^*+\omega_j)},~ i,j=1, 2, \ldots, n, \quad
\eea\label{nsol}\ees
$a_{\ell}$, $\psi_j$ and $\phi$ are block-matrices of dimensions ($1\times M$), ($M \times 1$) and ($n \times 1$), respectively, with elements $a_{\ell} = -\left(\alpha_1^{(\ell)}, \alpha_2^{(\ell)}, \ldots, \alpha_{n}^{(\ell)}\right)$, $\psi_j=\left(\alpha_j^{(1)},~\alpha_j^{(2)},\ldots,\alpha_j^{(M)}\right)^T$ and $\phi = \left(e^{\eta_1},e^{\eta_2},\ldots,e^{\eta_{n}}\right)^T$, where $\eta_j=k_j x-(ik_j^2+\omega_j)y+\omega_jt$, $j=1, 2, \ldots,n$, $\ell=1,2,3,...,M$. Here $k_j$, $\omega_j$ and $\alpha_{j}^{(\ell)}$,  $j=1,2,\ldots,n$, $\ell=1,2,\ldots,M$, are arbitrary complex parameters. The symbols $\dagger$ and $T$ appearing in the superscript indicate the transpose conjugate and transpose of the matrix, respectively, while $M$ and $n$ represent the component number and soliton number, respectively. The proof for the above bright $n$-soliton solution (\ref{nsol}) can be done by verifying that the bilinear equations (\ref{beq}) satisfy the Jacobi identity \cite{Kanna2009jpa}. One can also ascertain the integrability of the system by the existence of $n$-soliton solution, with arbitrary $n$.

\subsection{Bright one-soliton solution}
Here, we write the explicit form of bright one-soliton solution of $M$-LSRI system (\ref{model}), resulting for the choice $n=1$ in Eq. (\ref{nsol}), as below:
\bes\bea
&&S^{(\ell)}=  A_{\ell} \sqrt{k_{1R}\omega_{1R}}~\mbox{sech} \left(\eta_{1R}+\frac{R}{2}\right) e^{i(\eta_{1I}-\frac{\pi}{2})}, \quad \ell=1,2,...M,\\
&&L=-2k_{1R}^2 \mbox{sech}^2 \left(\eta_{1R}+\frac{R}{2}\right),
\eea\label{1sol}\ees
where $A_{\ell}={\alpha_1^{(\ell)}}{\left(\sum_{\ell=1}^{M} |\alpha_1^{(\ell)}|^2\right)^{-\frac{1}{2}}}$, $e^R=\frac{-\sum_{\ell=1}^{M} |\alpha_1^{(\ell)}|^2}{4k_{1R}\omega_{1R}}$, $\eta_{1R}=k_{1R}x+(2k_{1R}k_{1I}-\omega_{1R})y+\omega_{1R}t$ and $\eta_{1I}=k_{1I}x-(k_{1R}^2-k_{1I}^2+\omega_{1I})y+\omega_{1I}t$. In Eq. (\ref{1sol}), the subscript $R$ ($I$) appearing in a particular complex parameter denotes the real (imaginary) part of that complex parameter. The above bright one-soliton solution is characterized by ($M+2$) arbitrary complex parameters ($\alpha_1^{(\ell)},~ \ell=1,2,...M$, $k_1$ and $\omega_1$) and it becomes singular (non-singular) for the choice $e^R<0$ ($e^R>0$). So, one can obtain the regular solitons when the condition $e^R>0$ is satisfied, which restricts one of the parameters among $k_{1R}$ and $\omega_{1R}$ to be negative while the other takes positive values.
\begin{figure}[h]
\centering\includegraphics[width=0.3\linewidth]{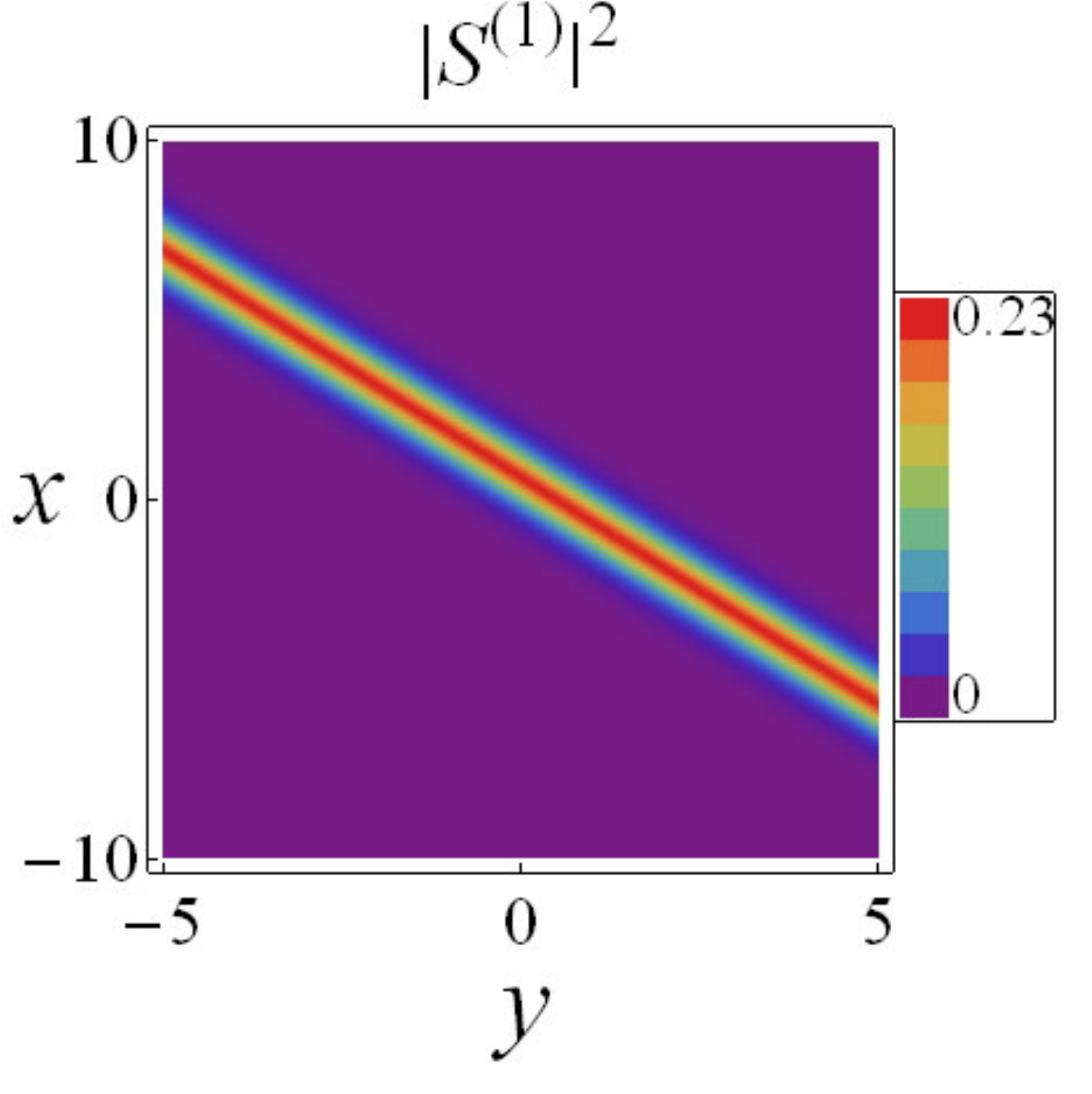}~~~\includegraphics[width=0.3\linewidth]{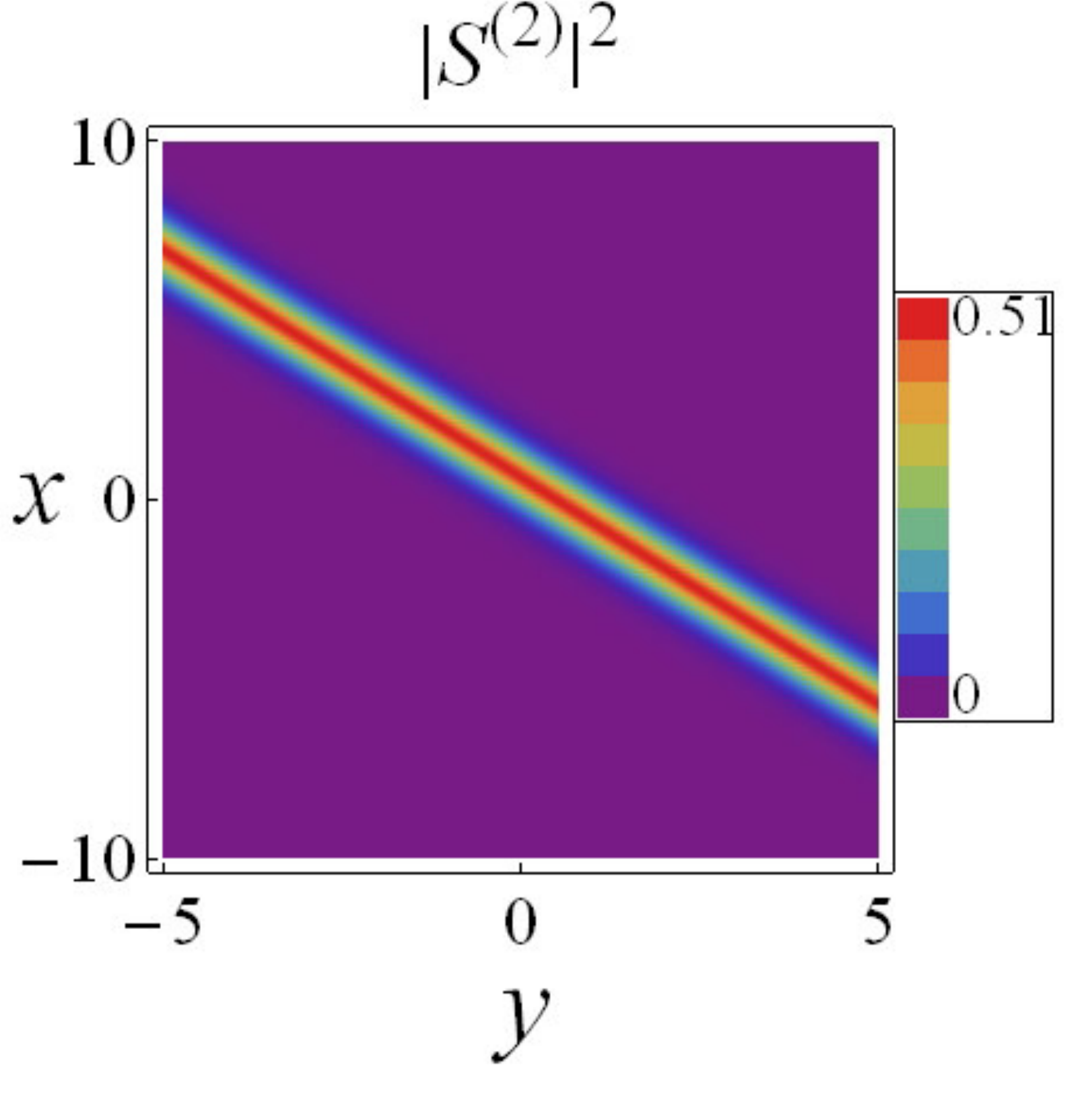}~~~\includegraphics[width=0.3\linewidth]{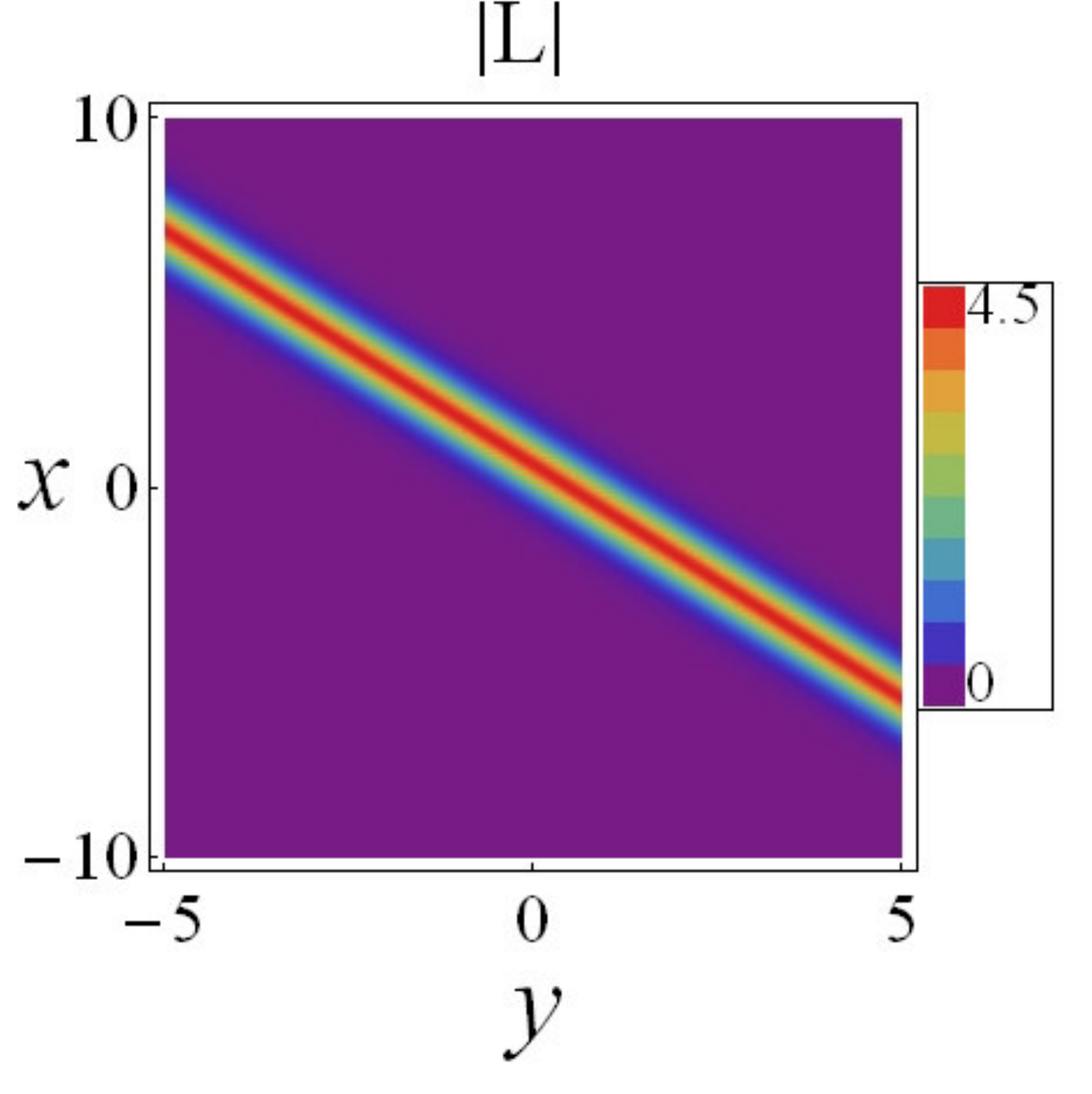}\\
\centering\includegraphics[width=0.3\linewidth]{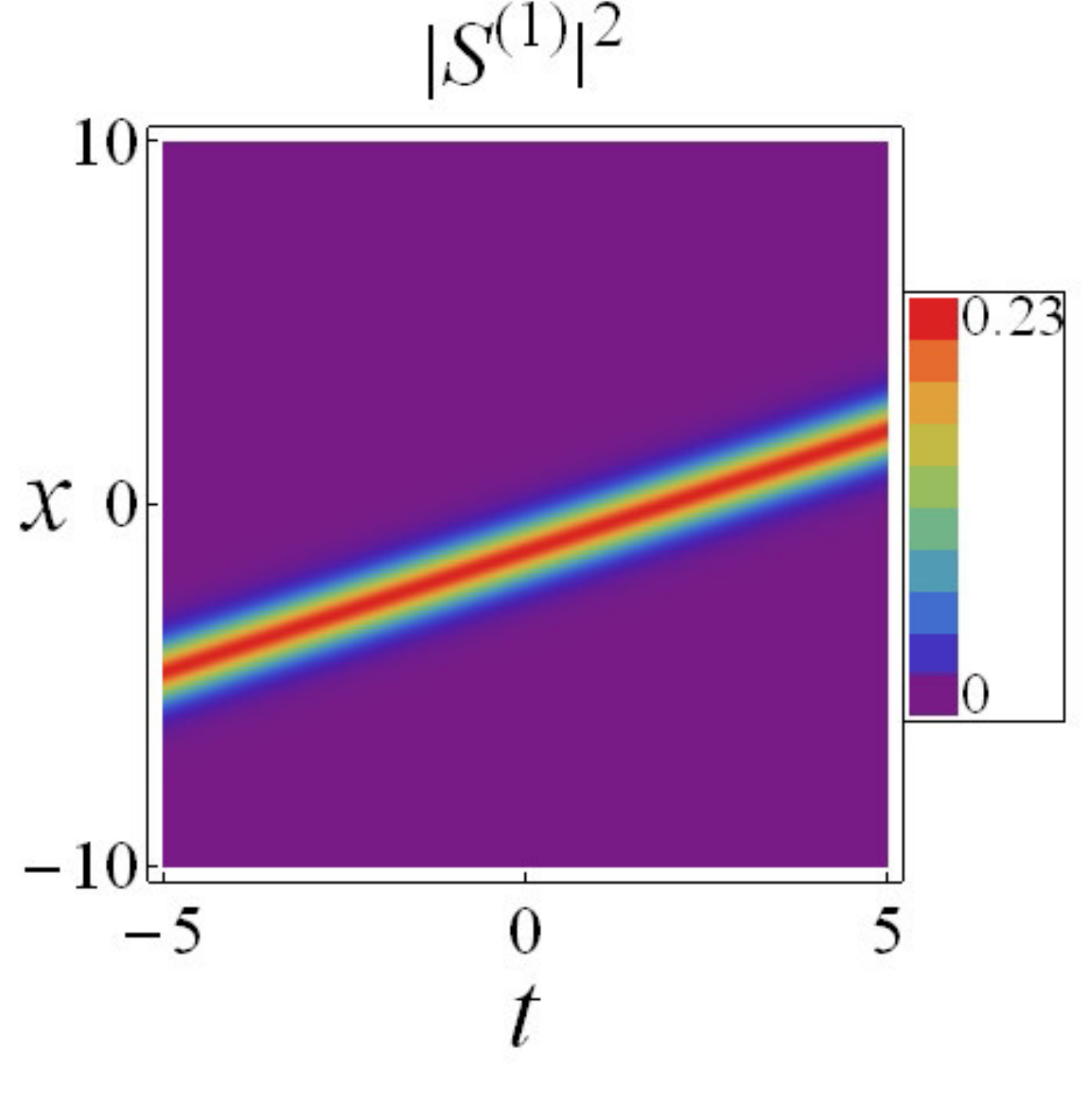}~~~\includegraphics[width=0.3\linewidth]{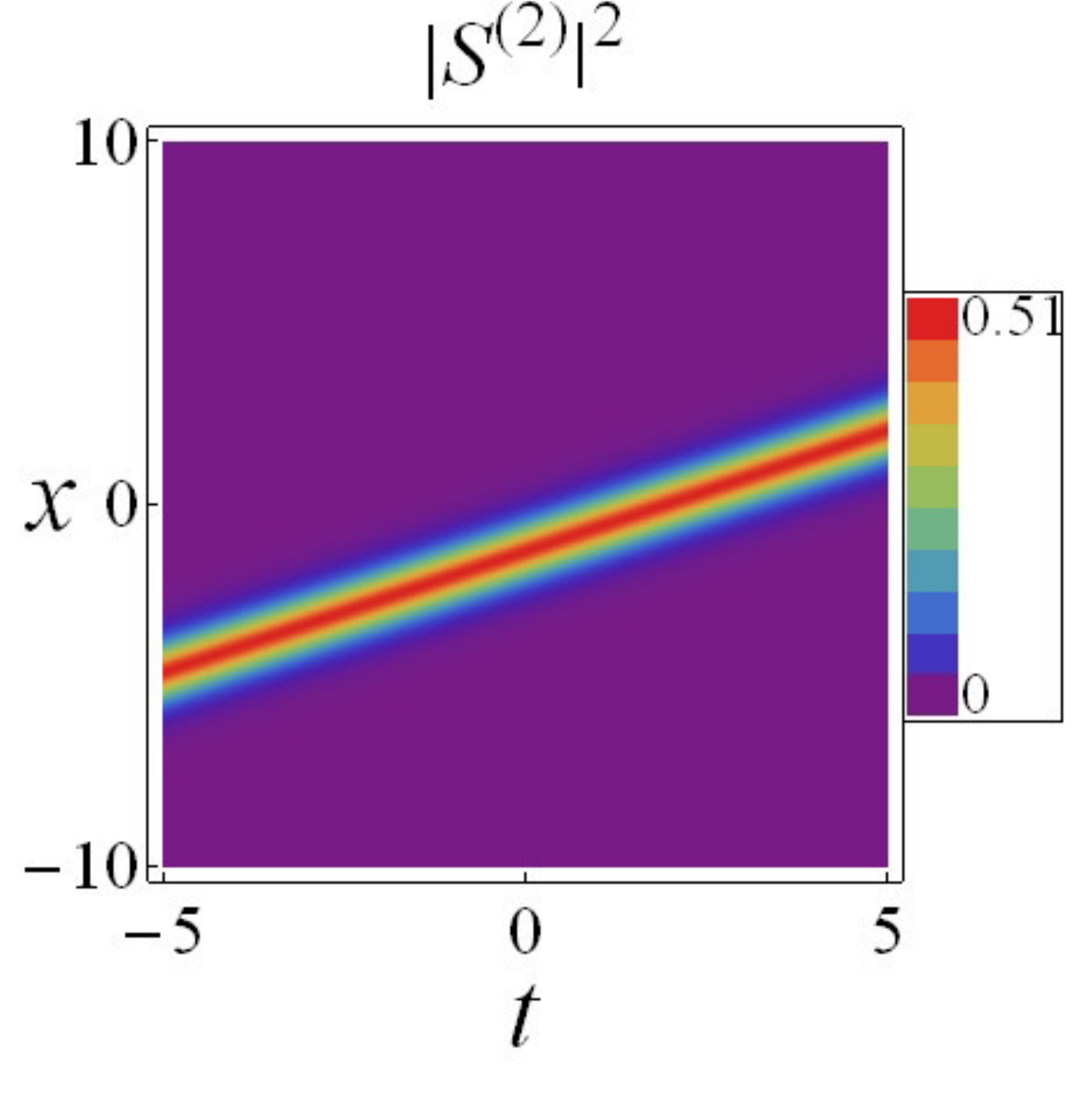}~~~\includegraphics[width=0.3\linewidth]{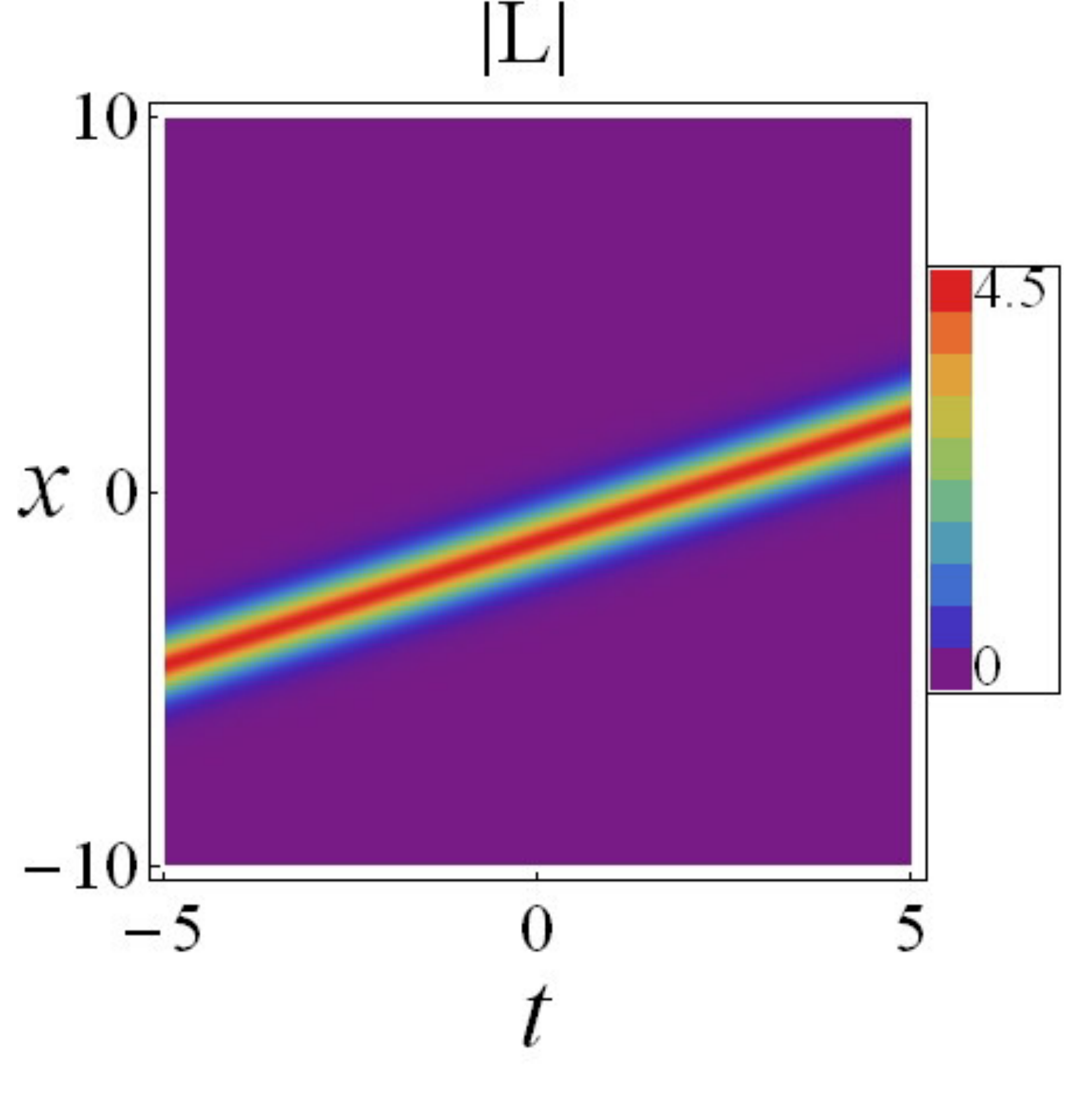}
\caption{Propagation of bright one-soliton of 2-LSRI system in the ($x-y$) plane for $t=1$ (top panels) and in the ($x-t$) plane for $y=1$ (bottom panels).}
\label{osfig1}
\end{figure}

The amplitude (peak value) of soliton in the LW component ($L$) is $2k_{1R}^2$ and that of the $\ell$th SW component ($S^{(\ell)},~\ell=1,2,...,M$) is $A_{\ell} \sqrt{k_{1R}\omega_{1R}}$. As the amplitude of soliton in the LW component is independent of $\alpha_1^{(\ell)}$ and $\omega_{1}$ parameters, one can control the soliton in the SW component by tuning these parameters without affecting the soliton in the LW component. Soliton of the present (2+1)D $M$-LSRI system can propagate in two planes, namely ($x-y$) plane and ($x-t$) plane with different velocities $(\frac{\omega_{1R}}{k_{1R}}-2k_{1I})$ and $-\frac{\omega_{1R}}{k_{1R}}$, respectively, for fixed $t$ and $y$. By tuning the $k_{1I}$ parameter one can alter the velocity of propagating bright soliton in the ($x-y$) plane without affecting the soliton velocity in the ($x-t$) plane. We have shown the propagation of bright one-soliton of 2-LSRI system in Fig. \ref{osfig1} for $k_1=1.5+0.3i$, $\omega_1=-1-2i$, $\alpha_1^{(1)}=1$ and $\alpha_1^{(2)}=1.5$.

\subsection{Bright two-soliton solution and their collisions}
Bright multi-solitons of the present system show interesting collision properties with energy sharing (energy-exchange or shape-changing) phenomenon, similar to the vector solitons in multicomponent Manakov system, coupled Gross-Pitaevskii equations, etc. \cite{RK1997pre,Kanna2001prl,Kannapramana,Kanna2003,Kanna2009epjst,Sakkara2013jmp,Kanna2014pla}. In order to understand this clearly, we consider the simple case of $n$-soliton solution, i.e., two-soliton solution [$n=2$ in Eq. (\ref{nsol})] of Eq. (\ref{model}) and analyze its dynamics. Since the solitons in the present (2+1)D $M$-LSRI system admit different velocities in the ($x-y$) and ($x-t$) planes, they show different collision characteristics in those planes. Particularly, the solitons can undergo both head-on and overtaking collisions in the ($x-y$) plane for different soliton parameters. Since the condition for non-singular solution restricts the velocity of solitons $\left(\frac{\omega_{1R}}{k_{1R}}~ \mbox{and }\frac{\omega_{2R}}{k_{2R}}\right)$ to be either positive or negative simultaneously, the solitons can undergo only overtaking collisions in the ($x-t$) plane. The bright solitons appearing in both components of 1-LSRI system (1SW and 1LW) exhibit only elastic collision. However, they undergo energy sharing collisions if there are two or more SW components, that is, $M$-LSRI system with $M\geq 2$.
\begin{figure}[h]
\centering\includegraphics[width=0.33\linewidth]{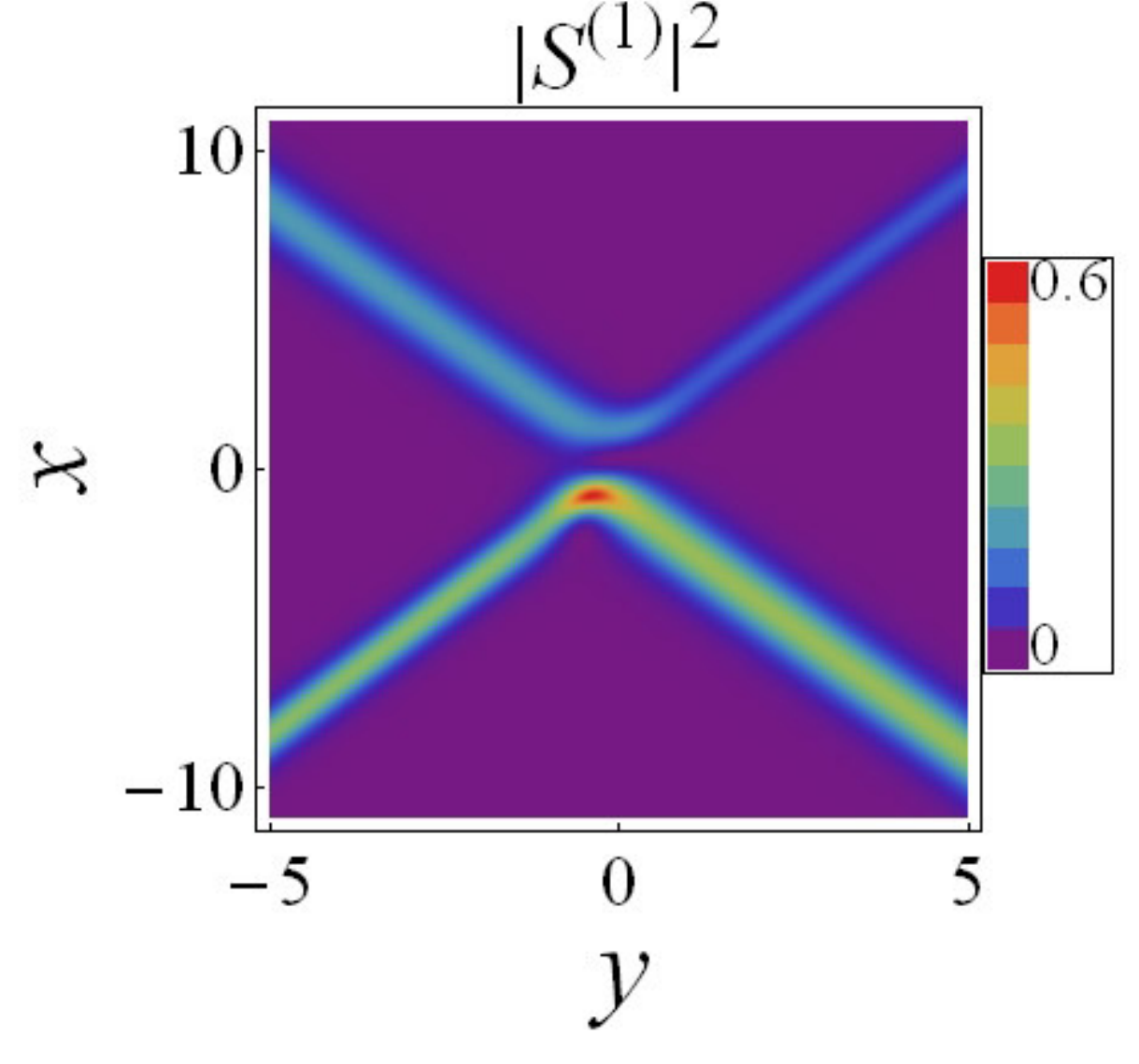}\includegraphics[width=0.33\linewidth]{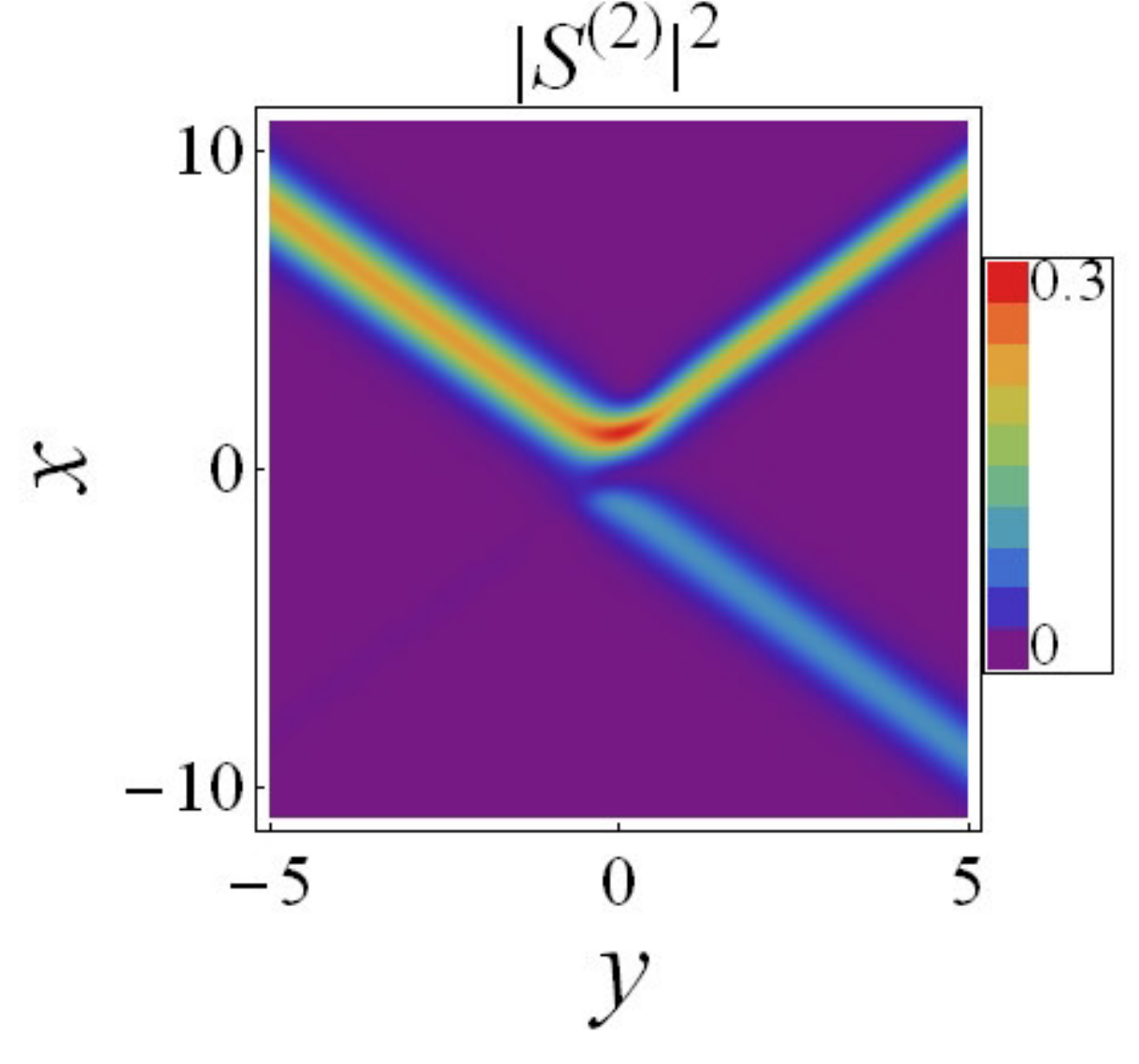}\includegraphics[width=0.33\linewidth]{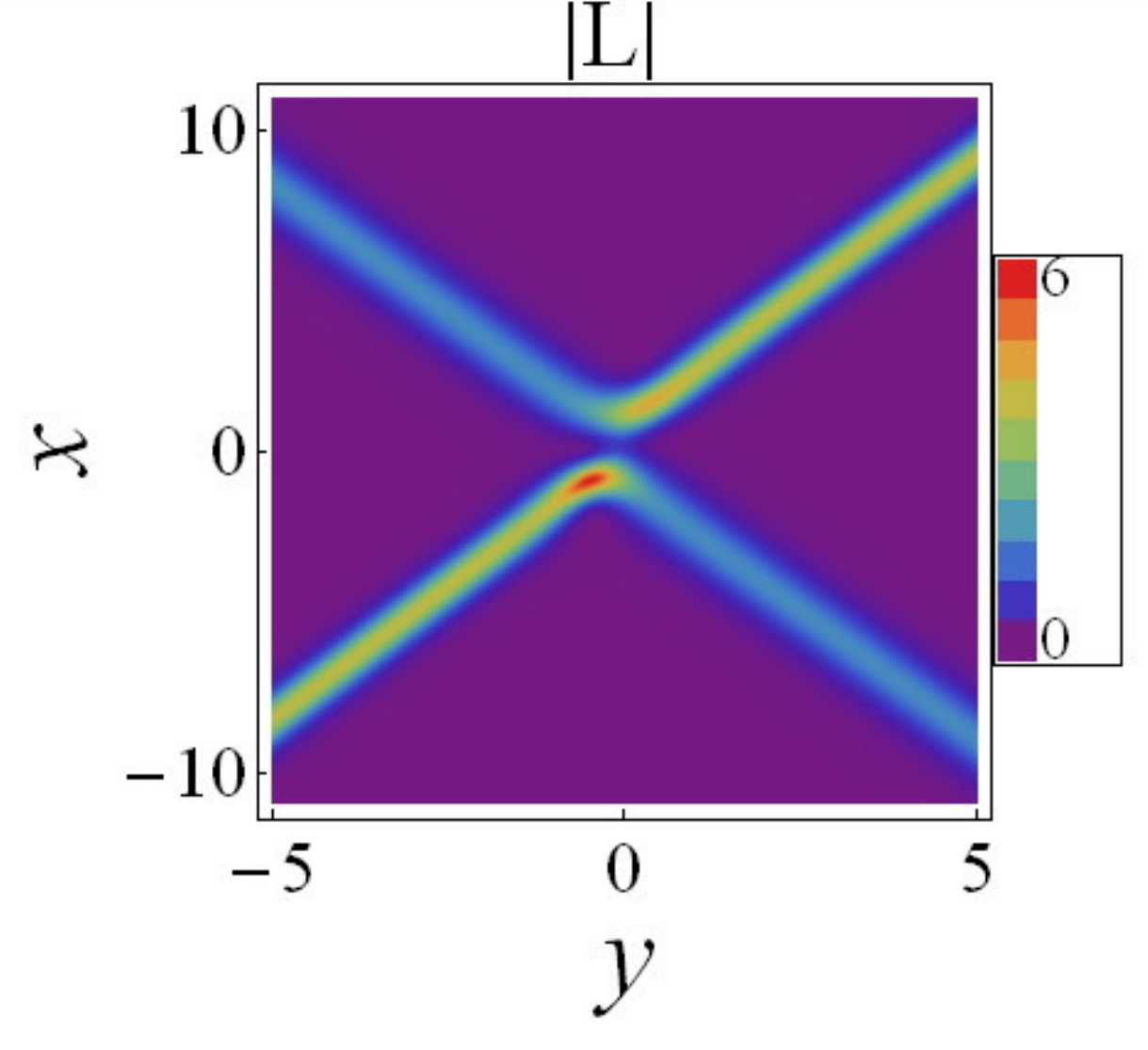}\\
\centering\includegraphics[width=0.33\linewidth]{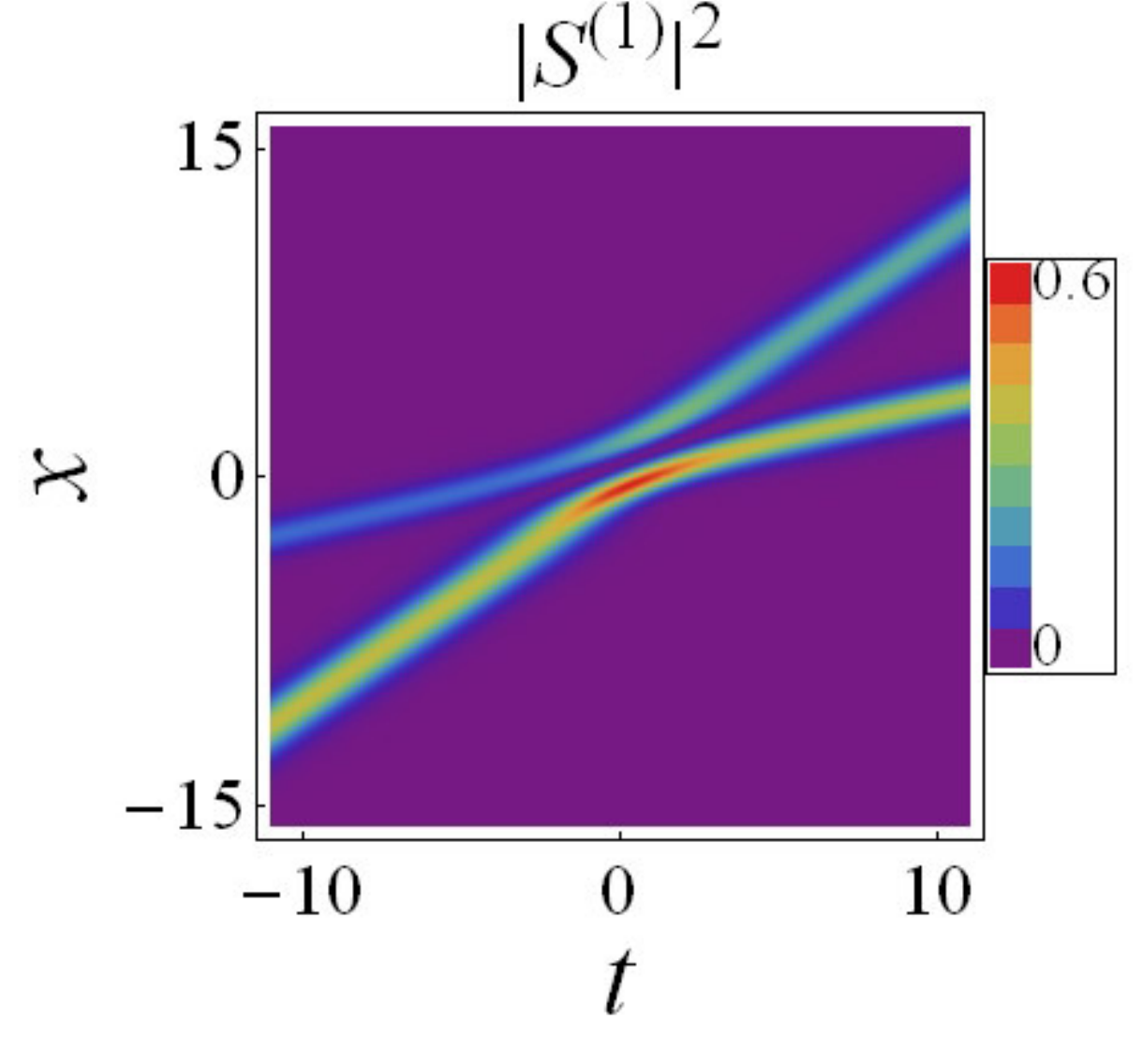}\includegraphics[width=0.33\linewidth]{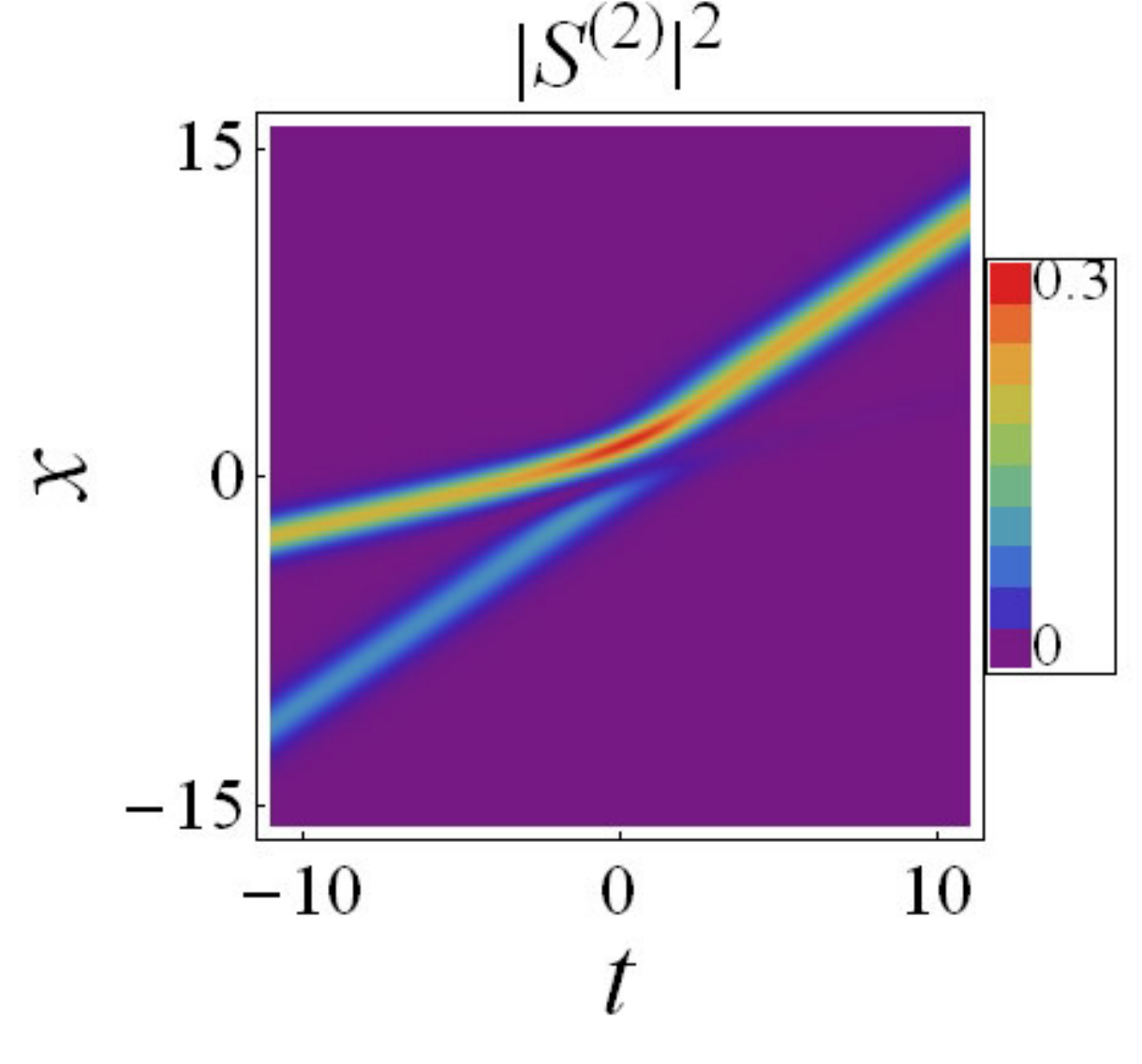}\includegraphics[width=0.33\linewidth]{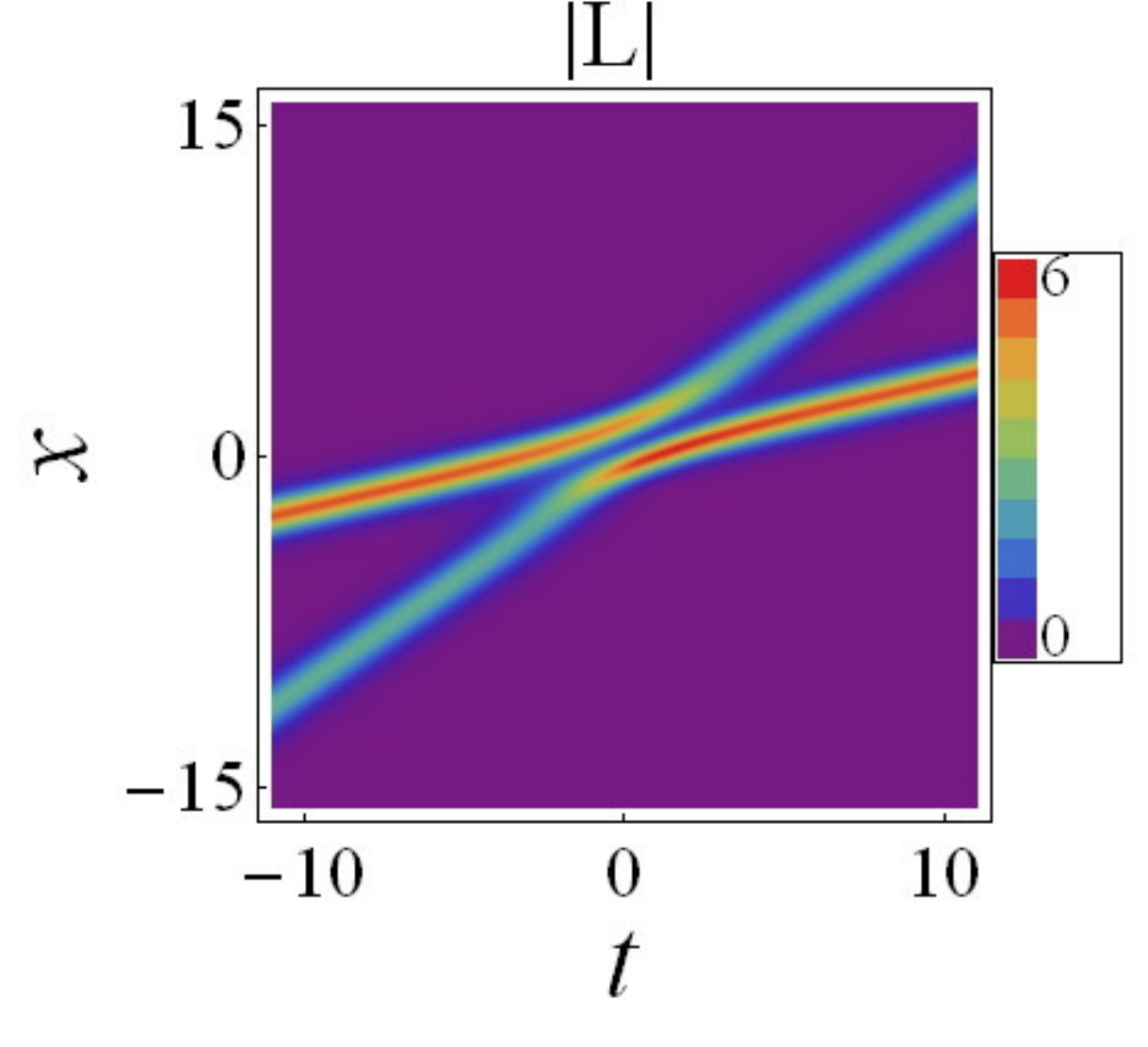}
\caption{Energy sharing collision of two bright solitons in 2-LSRI system. Head-on collision of solitons in the ($x-y$) plane at $t=1$ (top panels) and overtaking collision of solitons in the ($x-t$) plane at $y=1$ (bottom panels).}
\label{sc1fig1a}
\end{figure}

From a detailed asymptotic analysis \cite{Kanna2009jpa}, change in the amplitude of a given $j$-th soliton after collision in the $\ell$ -th SW component ($A_j^{(\ell)+}$) can be related to the amplitude of that soliton before collision ($A_j^{(\ell)-}$) in terms of the transition amplitudes ($T_j^{(\ell)}$) as
\bes\bea
A_j^{(\ell)+} = T_j^{(\ell)} A_j^{(\ell)-}, \qquad j=1,2, \qquad \ell=1,2,...,M,\label{asymp}
\eea
where 
\bea T_1^{(\ell)} &=& \frac{1-\lambda_1}{\sqrt{1-\lambda_1 \lambda_2}}\left(\frac{(k_1-k_2)(k_2+k_1^*)}{(k_1^*-k_2^*)(k_2^*+k_1)} \right)^{1/2},\\
T_2^{(\ell)} &=& \frac{\sqrt{1-\lambda_1 \lambda_2}}{1-\lambda_2} \left(\frac{(k_2+k_1^*)(k_1^*-k_2^*)}{(k_2^*+k_1)(k_1-k_2)} \right)^{1/2},
\eea \ees in which $\lambda_1=\frac{\alpha_2^{(\ell)} \kappa_{12}}{\alpha_1^{(\ell)} \kappa_{22}}$ and $\lambda_2=\frac{\alpha_1^{(\ell)} \kappa_{21}}{\alpha_2^{(\ell)} \kappa_{11}}$, where the form of $\kappa_{ij},~i,j=1,2$, are as given in Eq. (\ref{nsol}c) for $n=2$. The solitons undergo elastic collision for a special choice of soliton parameters ($\alpha_j^{(\ell)},~j=1,2,~\ell=1,2,...,M$) satisfying the condition $\frac{\alpha_1^{(1)}}{\alpha_2^{(1)}} = \frac{\alpha_1^{(2)}}{\alpha_2^{(2)}}= ... =\frac{\alpha_1^{(M)}}{\alpha_2^{(M)}}$, for which $T_j^{(\ell)}$ become uni-modular, i.e., $|T_j^{(\ell)}|^2=1$. However, the solitons appearing in the LW component undergo only elastic collision for all the choices of $\alpha_j^{(\ell)}$ parameter. Additionally, the soliton (say $s_j,~j=1,2$) appearing in all the components experiences a phase-shift ($\Phi_j$) given by $\Phi_1=\ln\left(\sqrt{1-\lambda_1 \lambda_2}\left|\frac{k_1-k_2}{k_1+k_2^*}\right| \right) \equiv -\Phi_2$.

The energy sharing collision scenario of two bright solitons is shown in Fig. \ref{sc1fig1a} for $k_1=1+0.3i$, $k_2=1.5-i$, $\omega_1=-1-i$, $\omega_2=-0.5-0.5i$, $\alpha_1^{(1)}=2$, $\alpha_2^{(1)}=1$, $\alpha_1^{(2)}=1$, $\alpha_2^{(2)}=0.08$. In the ($x-y$) plane, the amplitude of soliton $s_1$ ($s_2$) is enhanced (suppressed) in $S^{(1)}$ while the amplitude of soliton $s_1$ ($s_2$) gets suppressed (enhanced) in ($S^{(2)}$). The switching nature of soliton intensity (energy) in the ($x-t$) plane is opposite to the switching phenomenon in the ($x-y$) plane. However, in both ($x-y$) and ($x-t$) planes, the LW solitons emerge unaltered after collision except for a phase-shift.

\section{Dark solitons}\label{secdark}
As noted in the section \ref{secbilin}, dark soliton solutions of $M$-LSRI system (\ref{model}) result for the choice $\lambda \neq 0$ in the bilinear equations (\ref{beq}). In the following, we obtain the dark one- and two-soliton solutions of system (\ref{model}) by applying the Hirota's bilinearization method \cite{Kiv1993ol,RKjpadark,Ohtadark}.
\subsection{Dark one-soliton solution}
To construct the dark one-soliton solution, we choose the form of $g^{(\ell)})$ and $f$ as $g^{(\ell)}=g_0^{(\ell)}(1+\chi^2 g_2^{(\ell)})$, $\ell=1,2,...,M$, and $f=1+\chi^2 f_2$. By substituting these expressions into the bilinear equations (\ref{beq}) and recursively solving the resulting set of equations, we get the explicit expressions for $g^{(\ell)}$ and $f$ as $g^{(\ell)}=\tau_{\ell} (1+\mu_1^{(\ell)}e^{\eta_1})e^{i\psi_{\ell}}$, $\ell=1,2,...,M$, and $f=1+ e^{\eta_1}$.
Hence from Eq. (\ref{trans}), the dark one-soliton solution can be written as
\bes\bea
\hspace{-1.0cm}S^{(\ell)}&=& \frac{\tau_{\ell}}{2} \left[(1+\mu_1^{(\ell)})-(1-\mu_1^{(\ell)})\tanh(\eta_1/2)\right] e^{i\psi_{\ell}}, \quad \ell=1,2,...,M,\\
\hspace{-1.0cm}L&=& -\frac{k_1^2}{2}~\mbox{sech}^2(\eta_1/2).
\eea \label{1dsa} \ees
where $\eta_1=k_1 x+p_1 y+\omega_1 t$, $\psi_{\ell}=a_{\ell} x+b_{\ell} y+c_{\ell} t$, $\ell=1,2,...,M$, $\lambda=\sum_{\ell=1}^M |\tau_{\ell}|^2$, and
$\mu_1^{(\ell)}=\frac{2a_{\ell}k_1-p_1-\omega_1+ik_1^2}{2a_{\ell}k_1-p_1-\omega_1-ik_1^2}$, $\ell=1,2,...,M$. Here $a_{\ell}$, $b_{\ell}$, $c_{\ell}$, $k_1$, $p_1$ and $\omega_1$ are real parameters, while $\tau_{\ell}$ are complex parameters and they should satisfy the relations
\bea && \frac{4 k_1^3}{\omega_1}\ds\sum_{\ell=1}^M \frac{|\tau_{\ell}|^2}{(2a_{\ell}k_1-p_1-\omega_1)^2+k_1^4}= 1,\nonumber\eea
and
\bea c_{\ell}=a_{\ell}^2-b_{\ell},\quad \ell=1,2,...,M.\nonumber\eea The dark one-soliton solution (\ref{1dsa}) is characterized by ($4M+2$) arbitrary real parameters.
\begin{figure}[h]
\centering\includegraphics[width=0.33\linewidth]{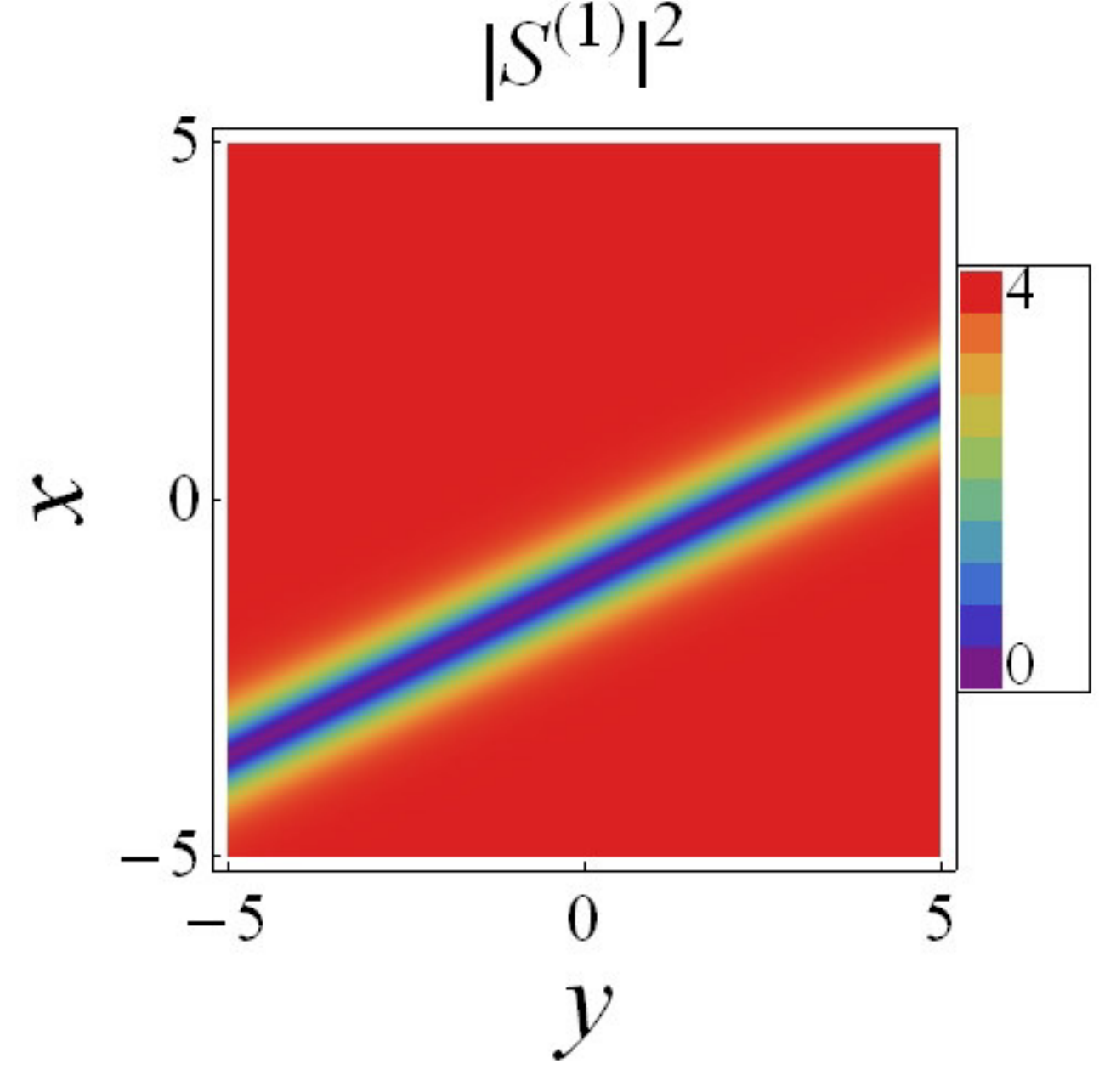}\includegraphics[width=0.33\linewidth]{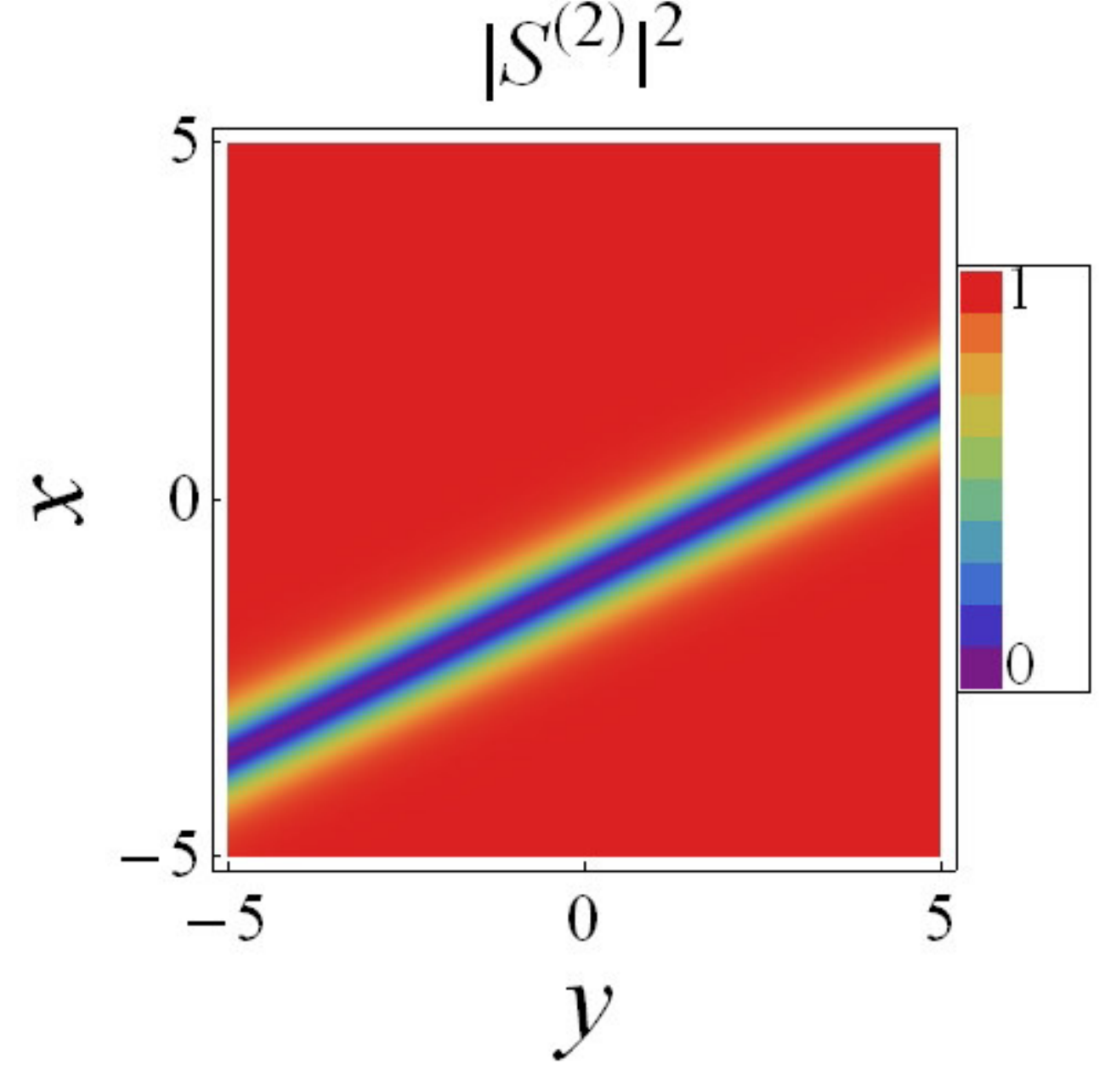}\includegraphics[width=0.33\linewidth]{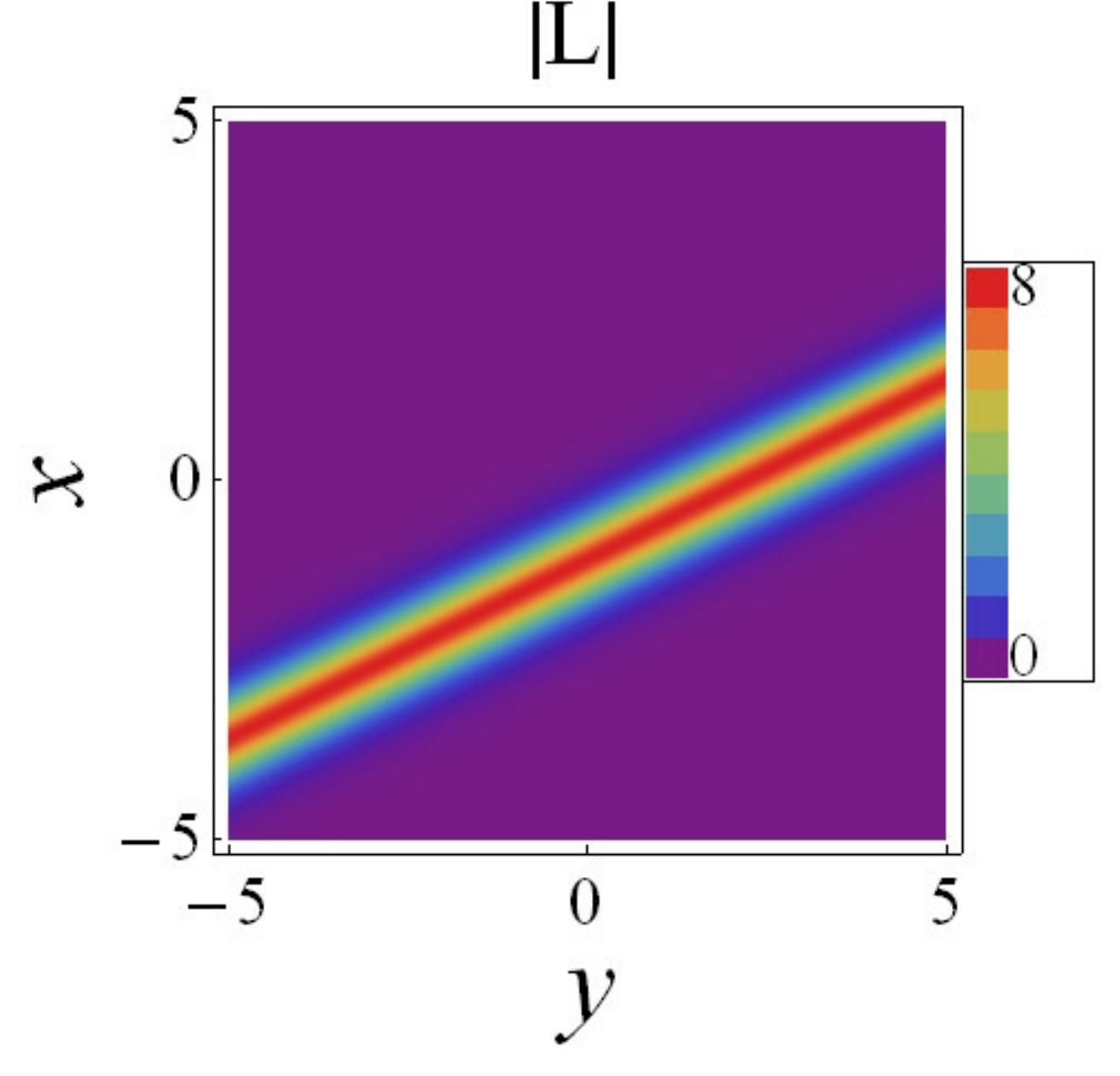}\\
\centering\includegraphics[width=0.33\linewidth]{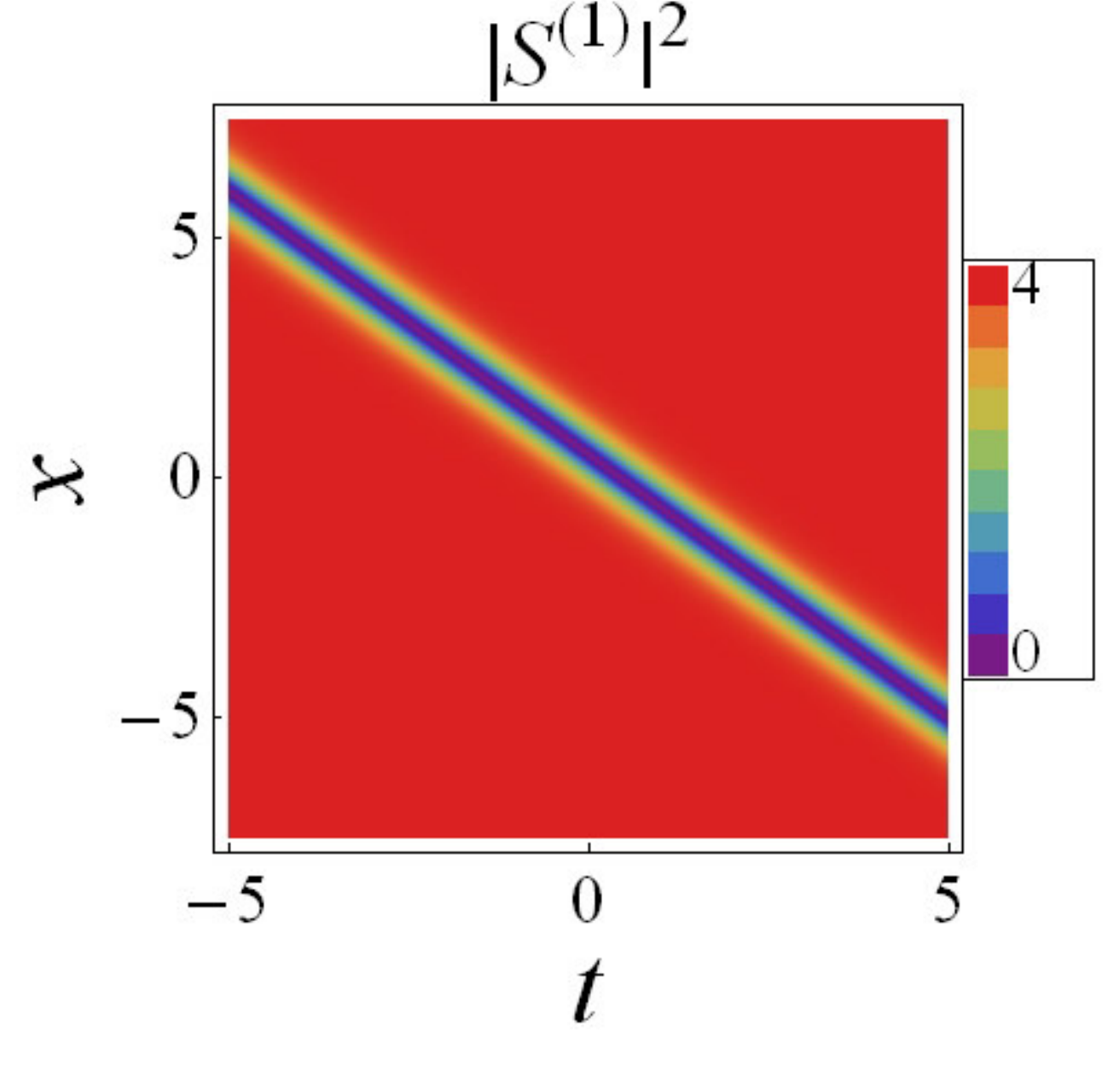}\includegraphics[width=0.33\linewidth]{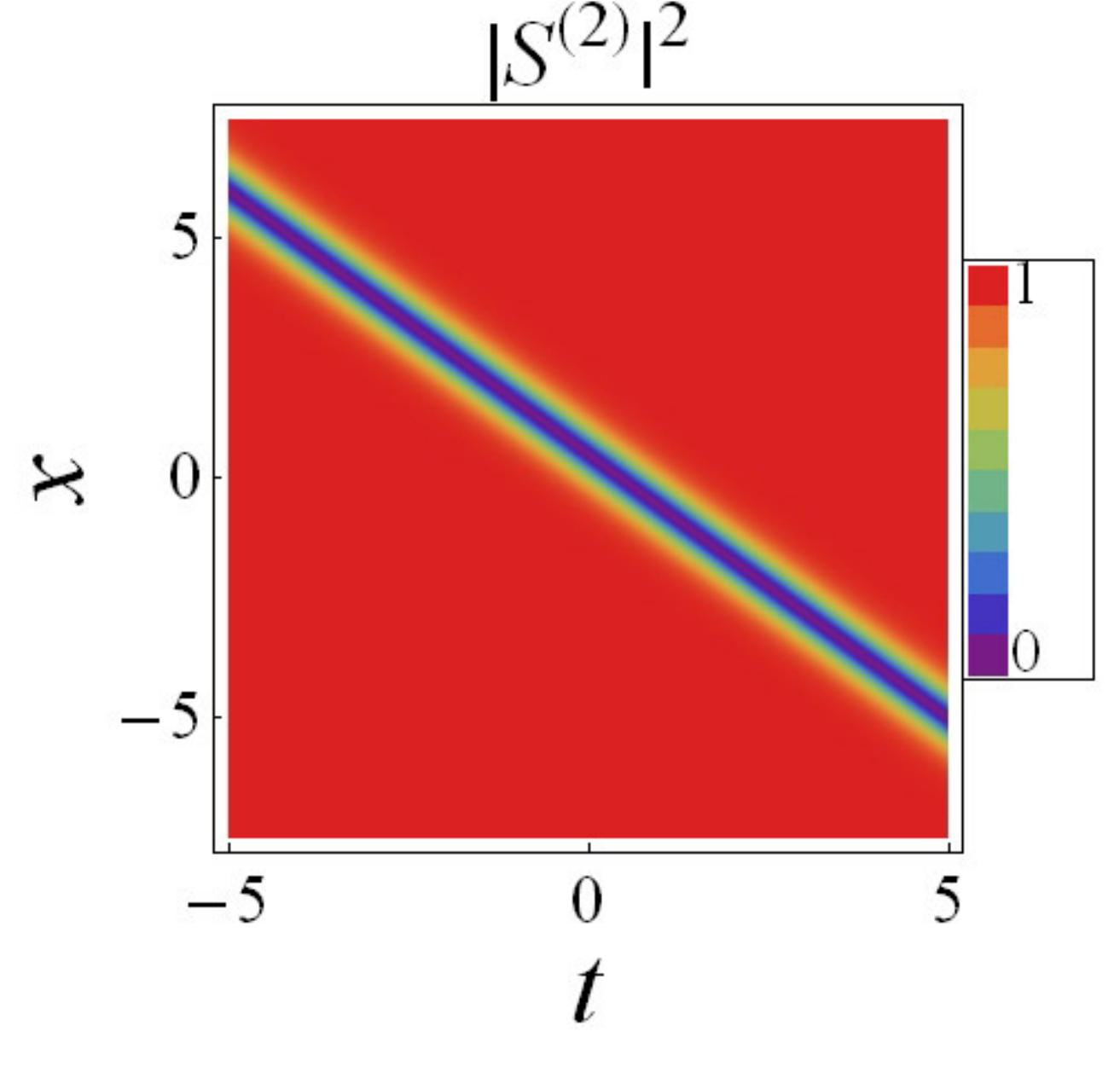}\includegraphics[width=0.33\linewidth]{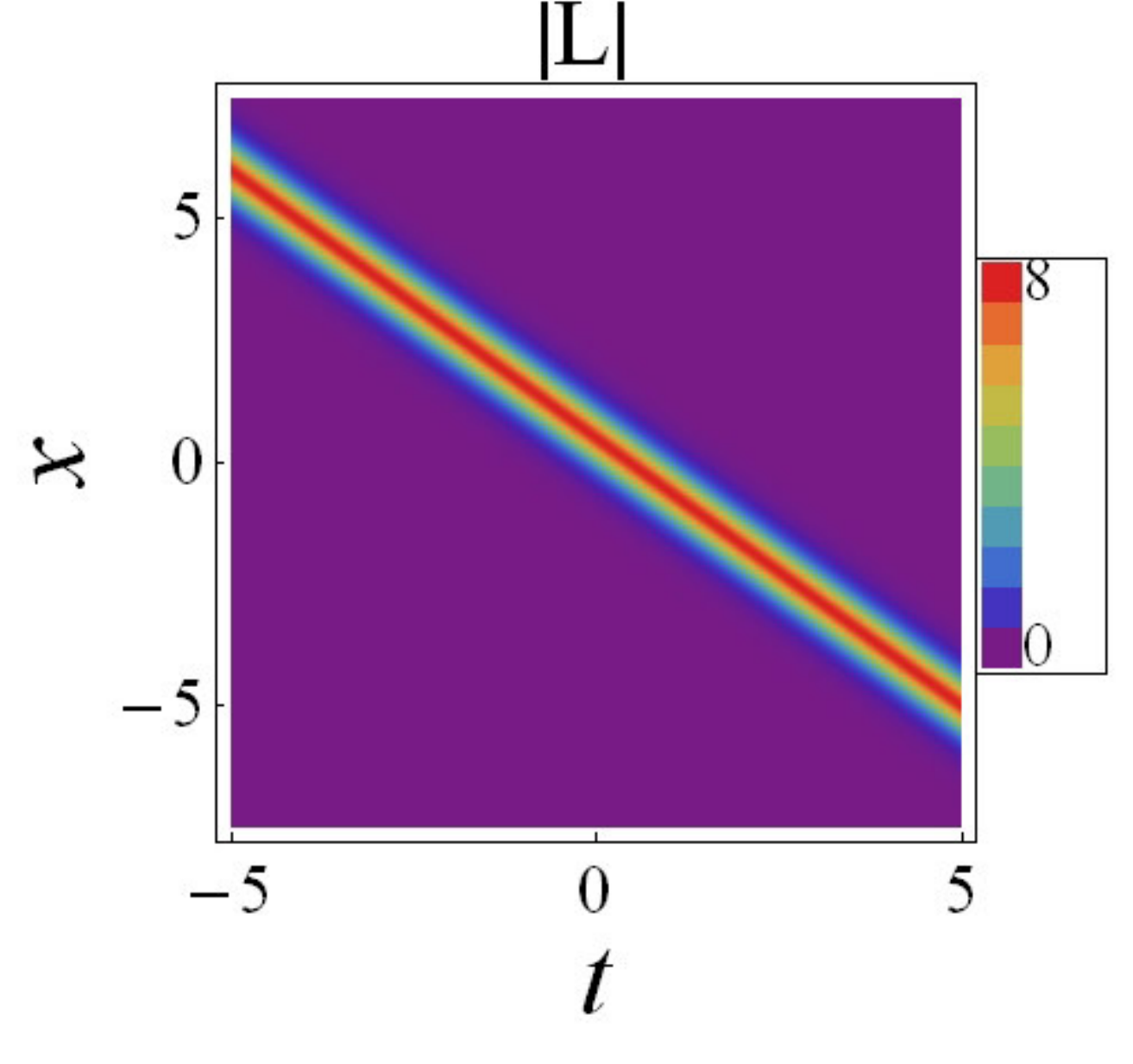}
\caption{Propagation of dark one-soliton of 2-LSRI system in ($x-y$) plane for $t=1$ (top panels) and in ($x-t$) plane for $y=1$ (bottom panels).}
\label{dsfig1s}
\end{figure}

The absolute square of the SW solution and the absolute of the LW solution given by the above equation (\ref{1dsa}) can be written in a compact form as
\bes \bea
|S^{(\ell)}|^2&=& |\tau_{\ell}|^2 \left[1-A_{\ell}~ \mbox{sech}^2(\eta_1/2)\right], \qquad \ell=1,2,...,M,\\
|L|&=& \frac{k_1^2}{2}~\mbox{sech}^2(\eta_1/2).
\eea \label{1dsb} \ees
where $A_{\ell}=\frac{k_1^4}{(2 a_{\ell} k_1-p_1-\omega_1)^2+k_1^4}$ determines the degree of darkness of dark soliton in the $\ell$-th SW component and $|\tau_{\ell}|^2$ represents its background intensity. Depending upon the values of $A_{\ell}$ parameters, one can get the dark and gray soliton in the SW components, that is, $A_{\ell}=1$ and $A_{\ell}<1$ result in dark and gray soliton, respectively. On the other hand, the LW component always results in the bright soliton with amplitude $\frac{k_1^2}{2}$, irrespective of other soliton parameters. The velocity of soliton (bright in LW and dark in SW) is $-\frac{p_1}{k_1}$ in the ($x-y$) plane and $-\frac{\omega_1}{k_1}$ in the ($x-t$) plane. So, the directions of soliton propagation (velocities) in both planes can be made different by controlling these quantities. We have shown a typical dark (bright) soliton propagation appearing in the SW (LW) component of 2-LSRI system in Fig. \ref{dsfig1s} for $k_1=4$, $p_1=-2$, $a_1=1$, $a_2=1.2$, $b_1=1.1$, $b_2=1.3$, $\tau_1=2$ and $\tau_2=1$.

\subsection{Dark two-soliton solution and their collision}
The dark two-soliton solution can be constructed by restricting the power series expansions for $g^{(\ell)}$ and $f$ as $g^{(\ell)}=g_0^{(\ell)}(1+\chi^2 g_2^{(\ell)}+\chi^4 g_4^{(\ell)})$, $\ell=1,2,...,M$, and $f=1+\chi^2 f_2+\chi^4 f_4$. The explicit forms of $g^{(\ell)}$ and $f$ can be obtained as
\bes\bea
\hspace{-1.0cm}g^{(\ell)}&=& \tau_{\ell} \left(1+\mu_1^{(\ell)}e^{\eta_1}+\mu_2^{(\ell)}e^{\eta_2}+\mu_1^{(\ell)}\mu_2^{(\ell)}\Omega e^{\eta_1+\eta_2}\right)e^{i\psi_{\ell}}, \quad \ell=1,2,...,M,\\
\hspace{-1.0cm}f&=&1+ e^{\eta_1}+ e^{\eta_2}+\Omega  e^{\eta_1+\eta_2},
\eea \label{2ds} \ees
where $\eta_j=k_j x+p_j y+\omega_j t$, $\psi_{\ell}=a_{\ell} x+b_{\ell} y+c_{\ell} t$, \bea\lambda=\sum_{\ell=1}^M |\tau_{\ell}|^2,\quad \mu_j^{(\ell)}=\frac{2a_{\ell}k_j-p_j-\omega_j+ik_j^2}{2a_{\ell}k_j-p_j-\omega_j-ik_j^2}, \quad j=1,2,~ \ell=1,2,...,M,\nonumber\eea and \bea \Omega =\frac{k_1^2 k_2^2(k_1-k_2)^2+(k_1(p_2+\omega_2)-k_2(p_1+\omega_1))^2}{k_1^2 k_2^2(k_1+k_2)^2+(k_1(p_2+\omega_2)-k_2(p_1+\omega_1))^2}\nonumber\eea. The above two-soliton solution is characterized by $(3M+6)$ real parameters $a_{\ell}$, $b_{\ell}$, $c_{\ell}$, $k_j$, $p_j$ and $\omega_j$, and $M$ complex parameters $\tau_{\ell}$, with $(M+2)$ relations \bea &&c_{\ell}=a_{\ell}^2-b_{\ell}, \quad \ell=1,2,...,M,\nonumber\\ 
&&\frac{2}{\omega_j k_j} \ds\sum_{\ell=1}^M |\tau_{\ell}|^2 \left(1-Re[\mu_j^{(\ell)}]\right)=1, \quad j=1,2.\nonumber\eea Hence we have only $(4M+4)$ number of arbitrary real constants. The velocities and darkness (amplitude) of dark (bright) solitons appearing in the SW (LW) components can be controlled by tuning these arbitrary parameters.
\begin{figure}[h]
\centering\includegraphics[width=0.33\linewidth]{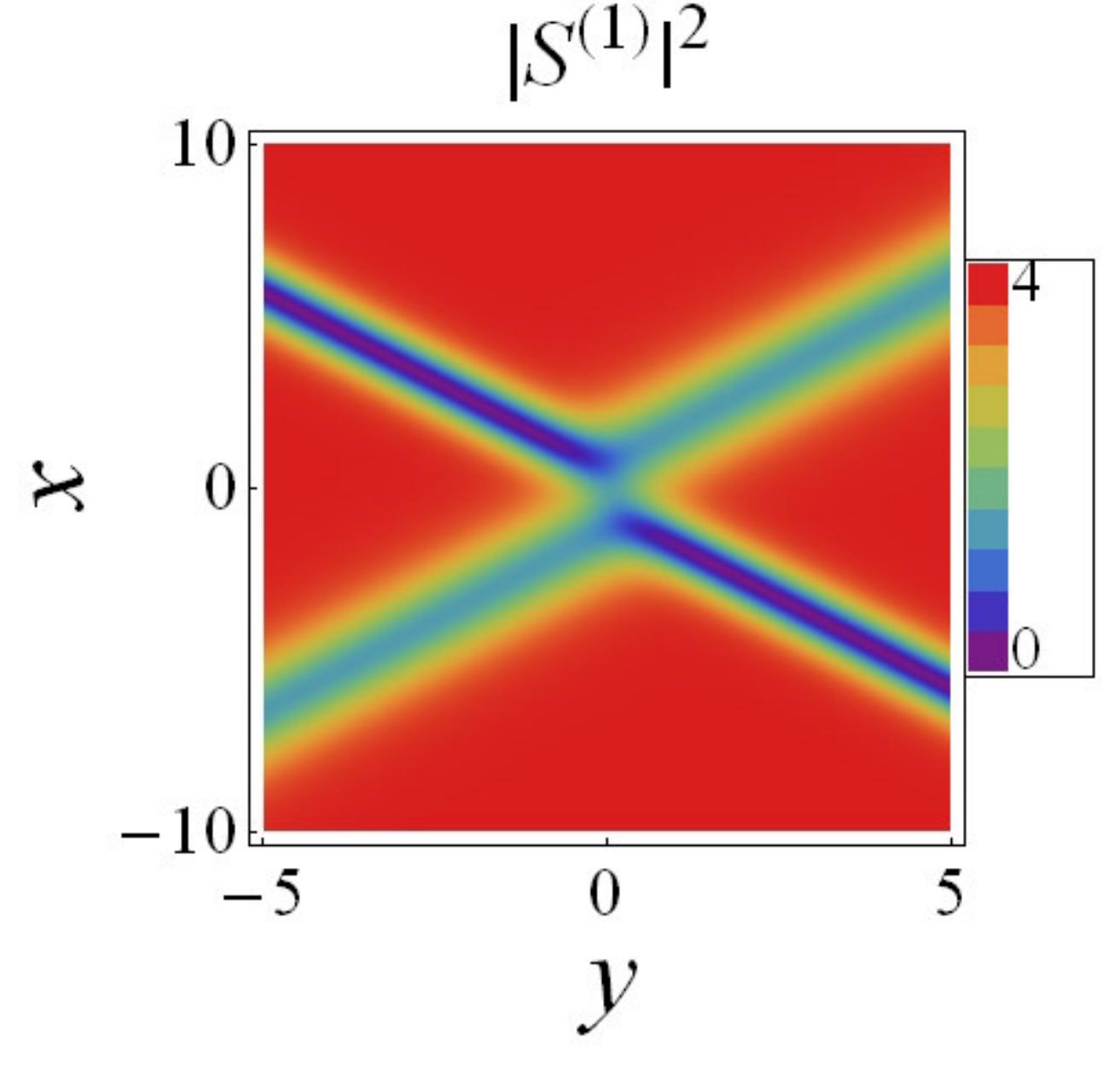}\includegraphics[width=0.33\linewidth]{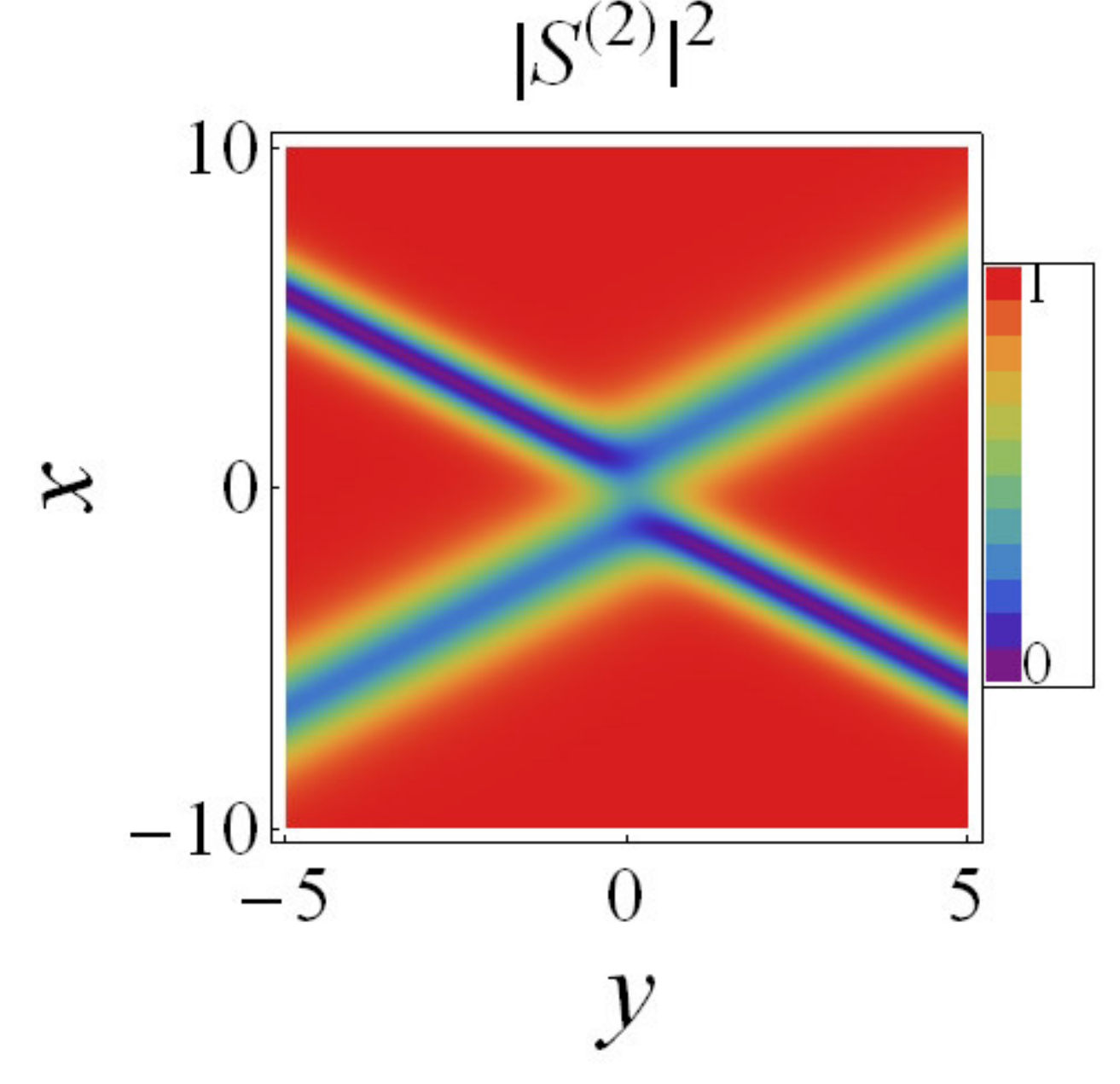}\includegraphics[width=0.33\linewidth]{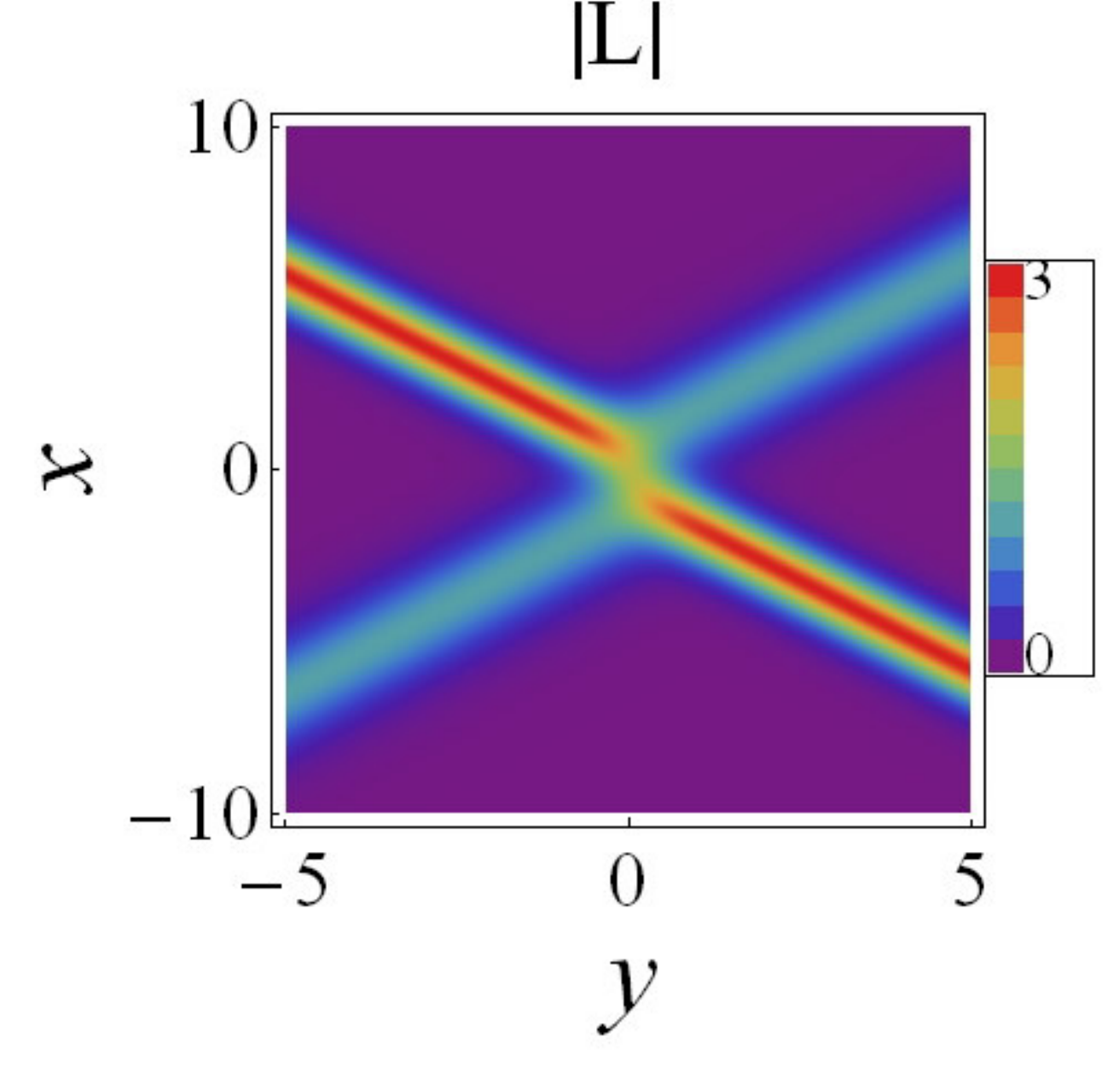}\\
\centering\includegraphics[width=0.33\linewidth]{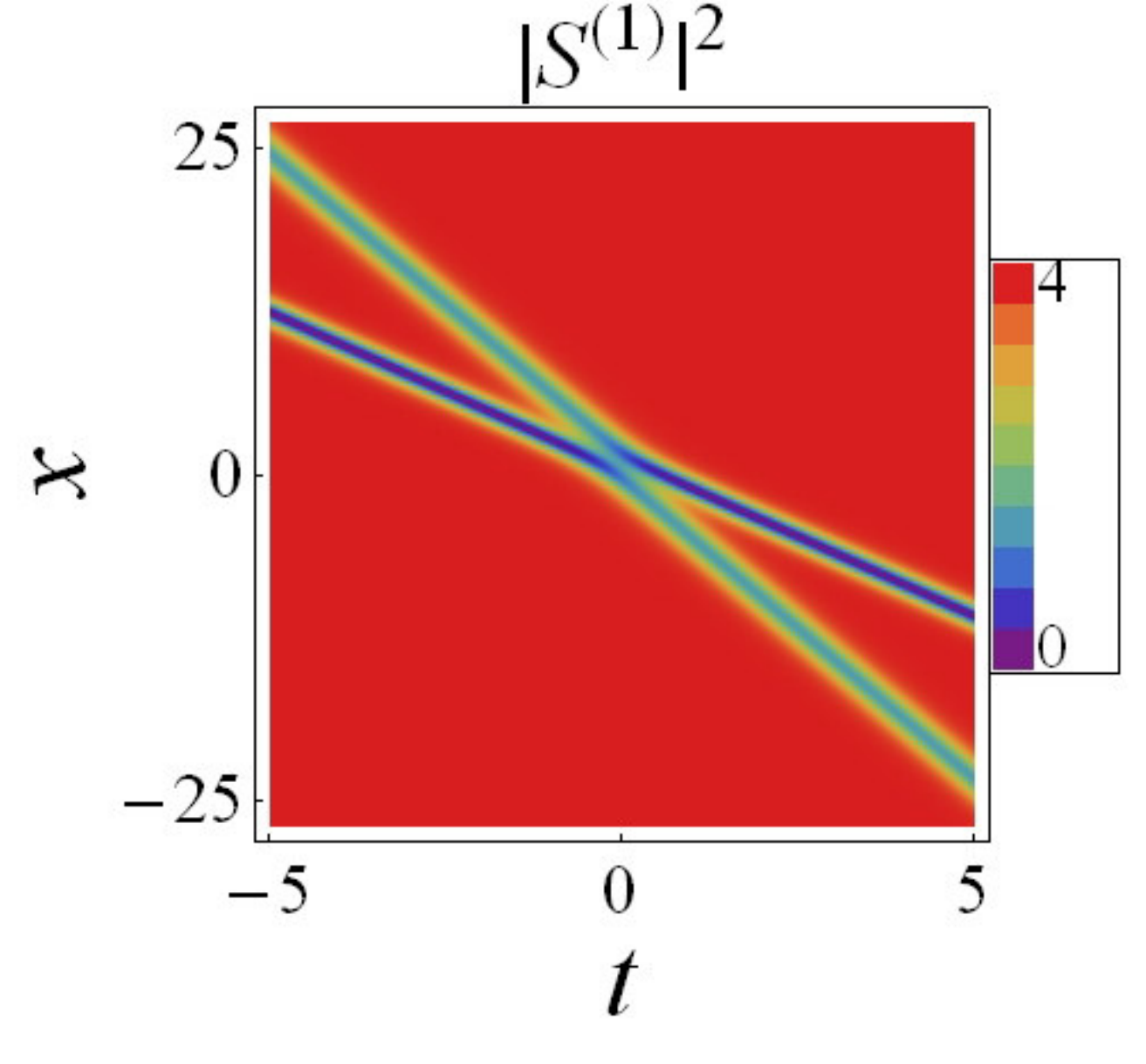}\includegraphics[width=0.33\linewidth]{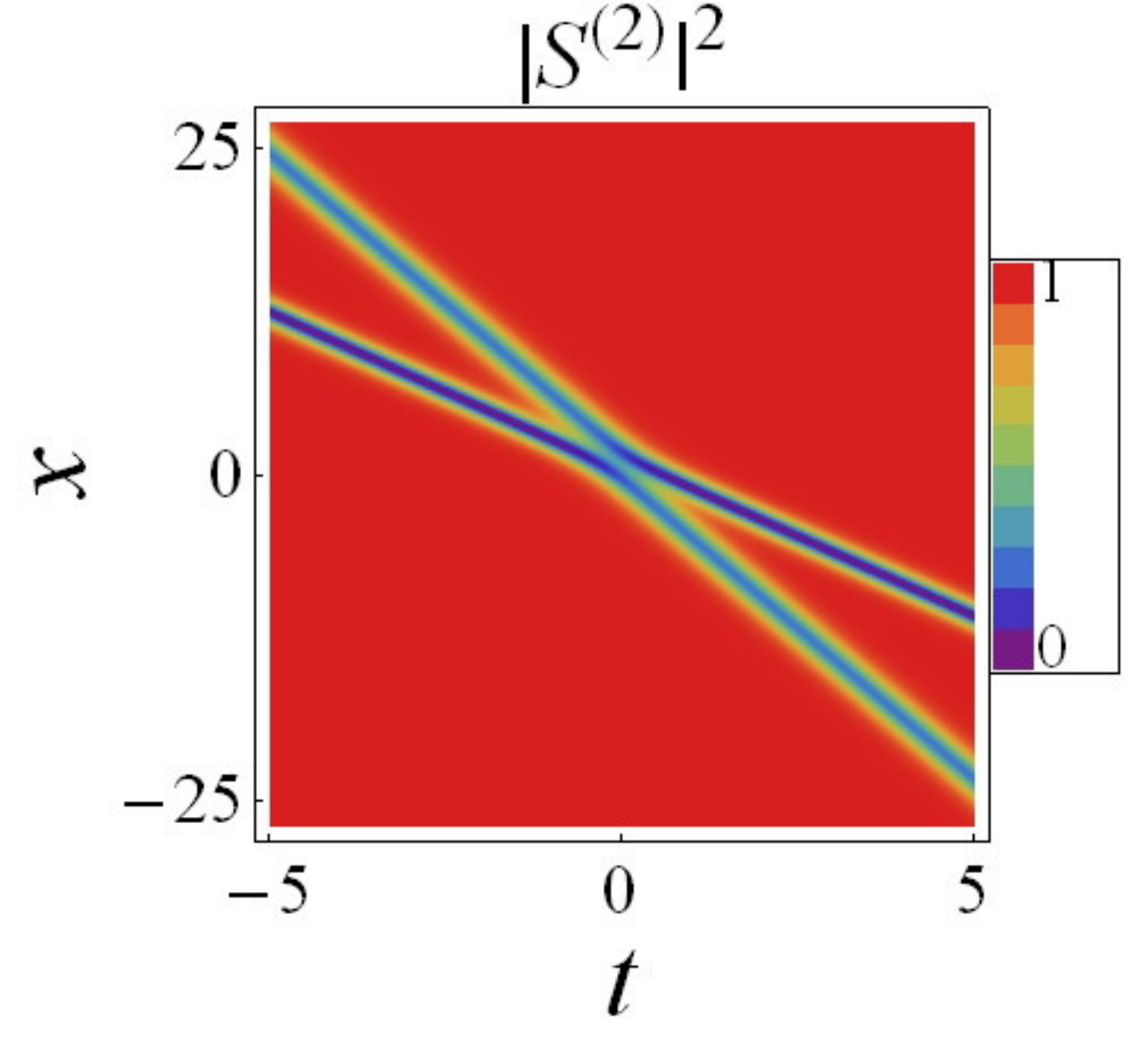}\includegraphics[width=0.33\linewidth]{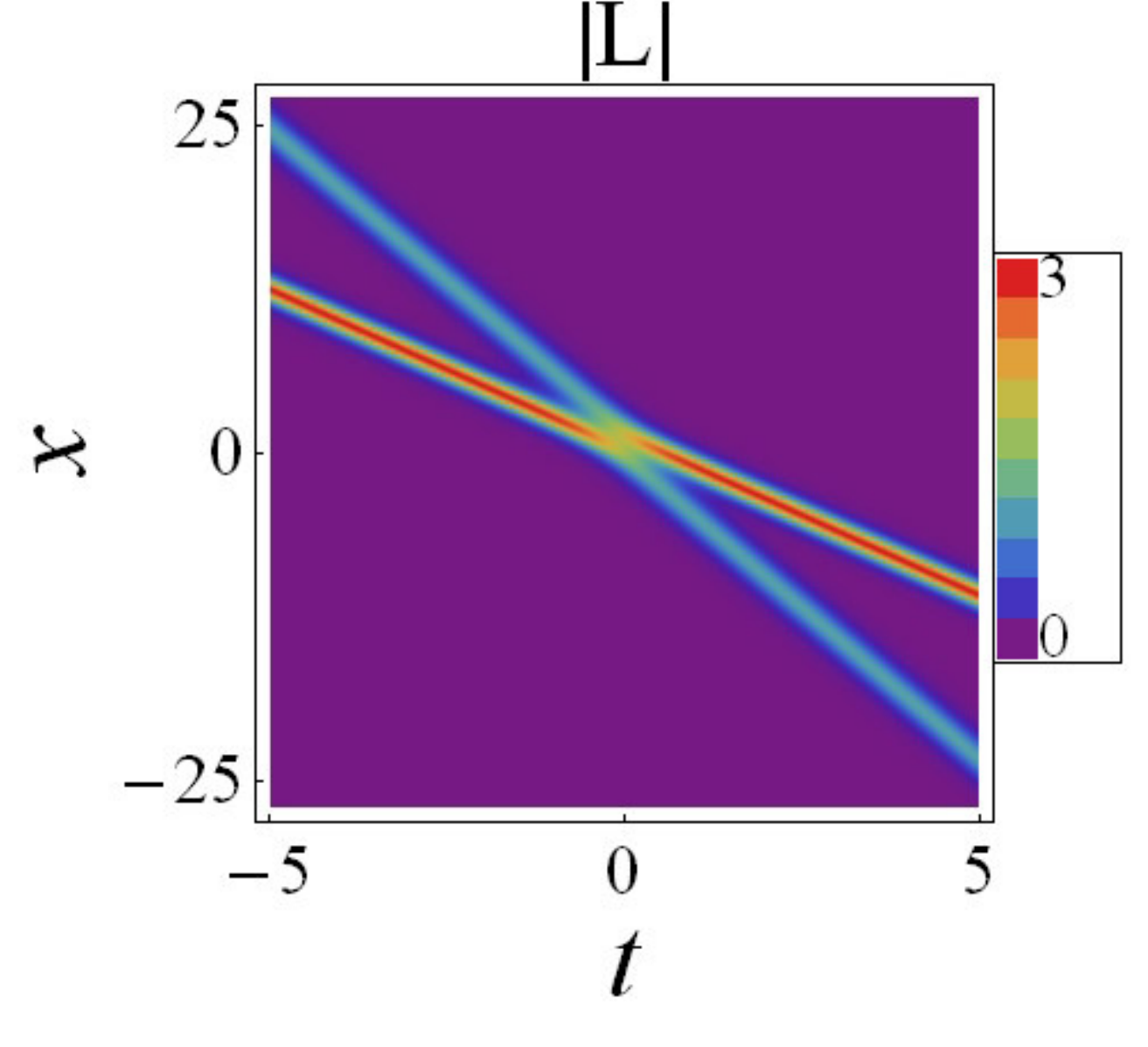}
\caption{Elastic collision of bright (dark-gray) solitons in the LW (SW) component(s) in 2-LSRI system in the ($x-y$) plane at $t=0.25$ (top panels) and in the ($x-t$) plane at $y=-0.25$ (bottom panels).}
\label{fig2ds}
\end{figure}

\begin{figure}[h]
\centering\includegraphics[width=0.33\linewidth]{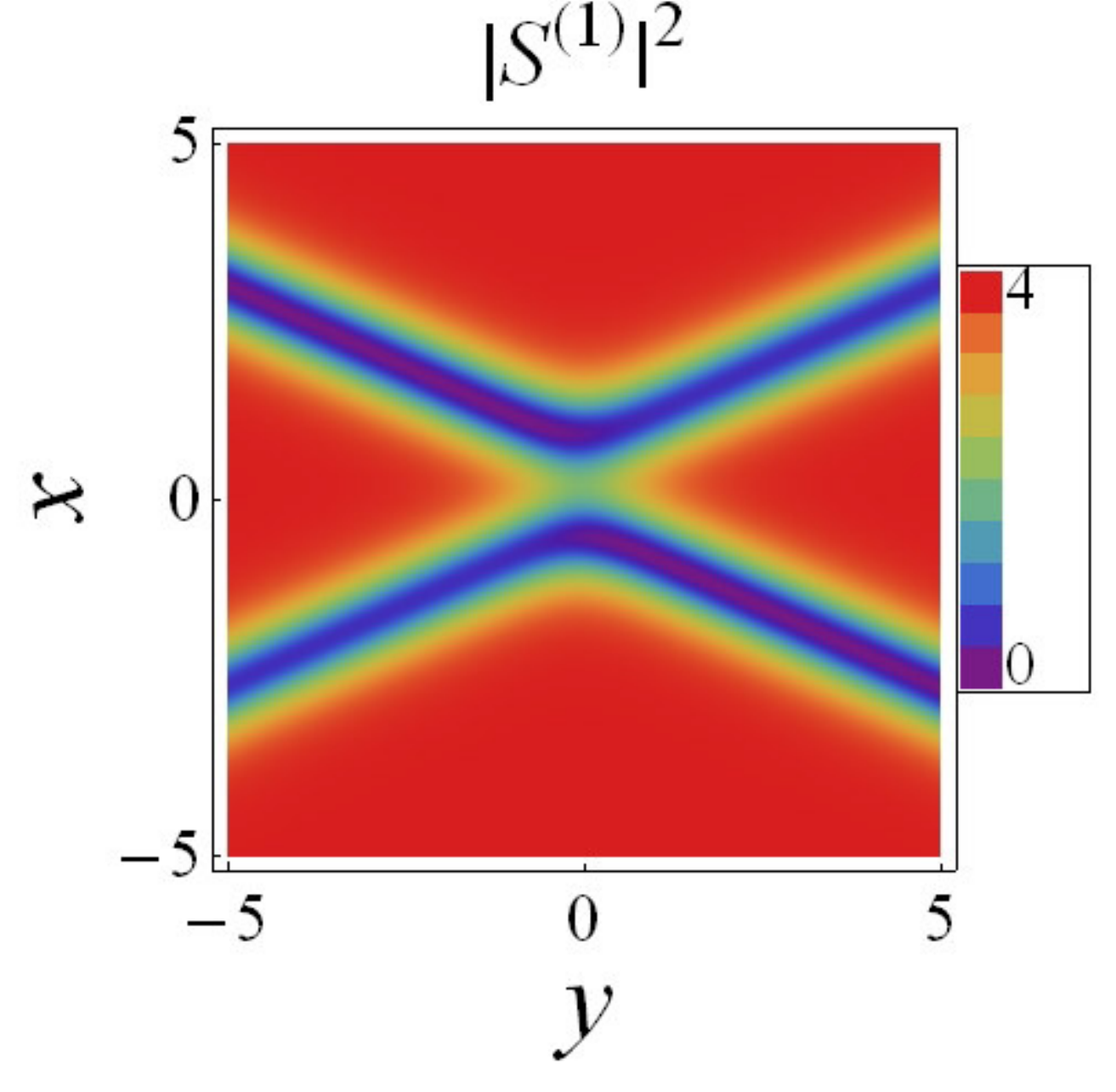}\includegraphics[width=0.33\linewidth]{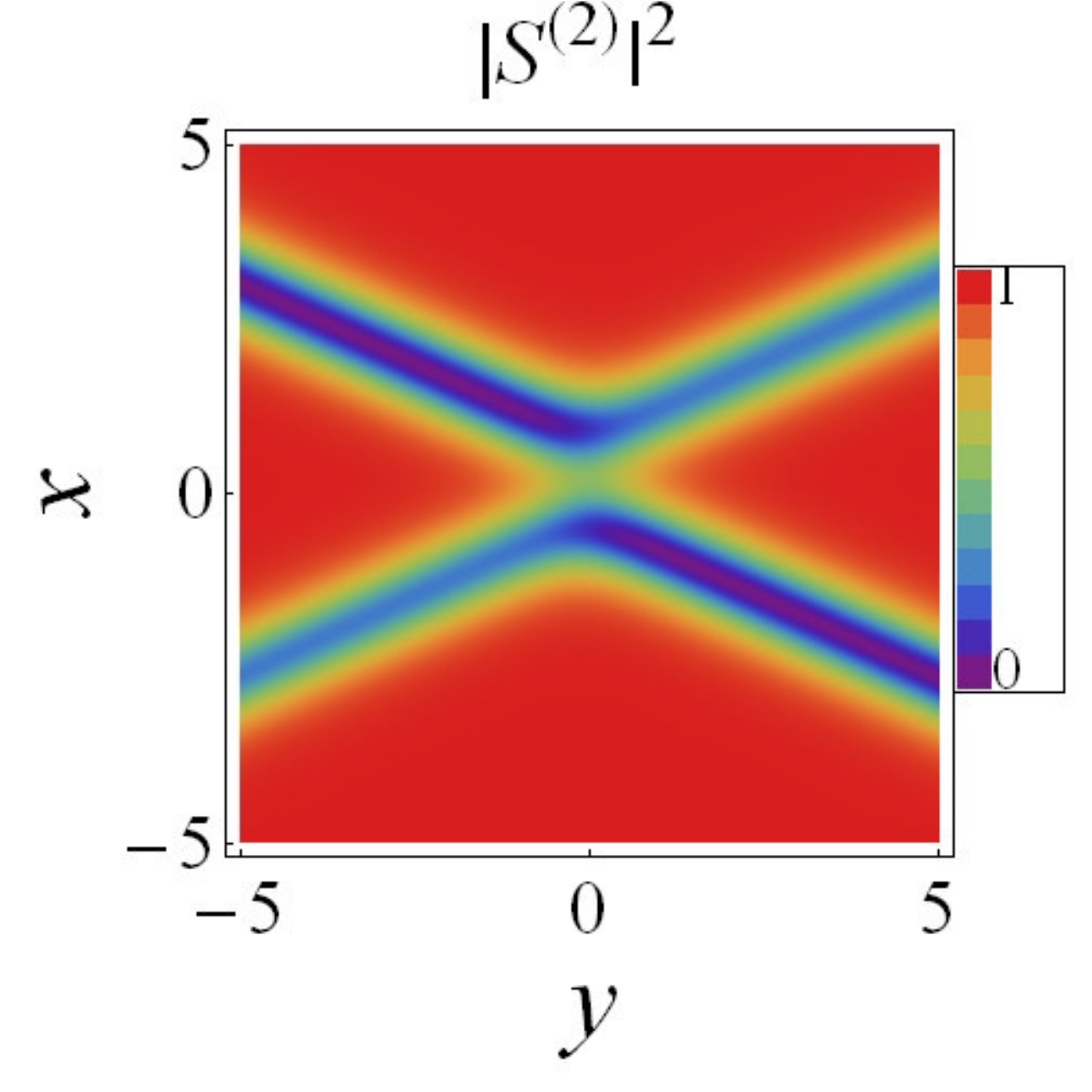}\includegraphics[width=0.33\linewidth]{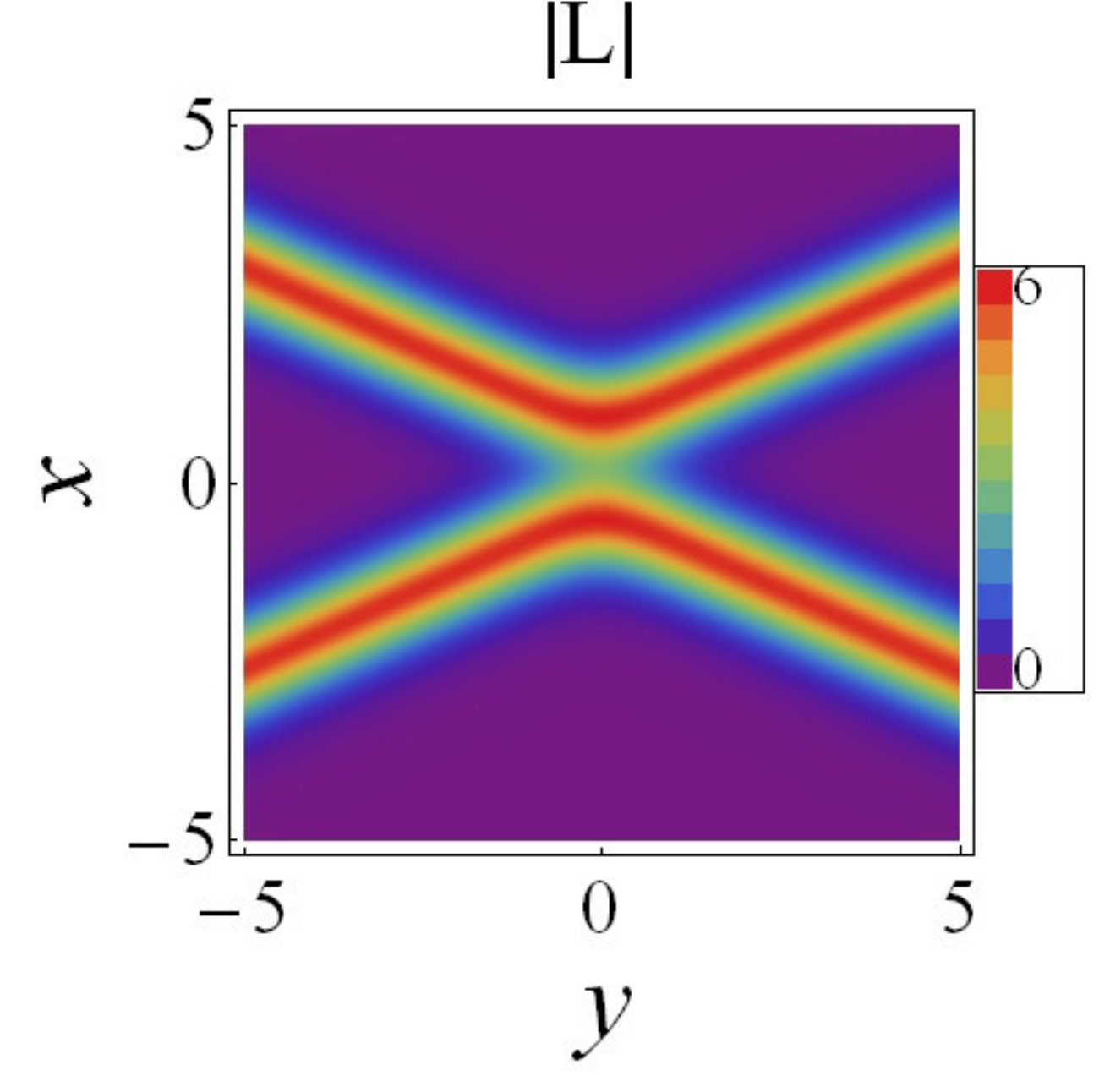}\\
\caption{Elastic collision of dark-dark, dark-gray and bright-bright solitons in the $S^{(1)}$, $S^{(2)}$ and $L$ components of 2-LSRI system in the ($x-y$) plane for $k_1=3.5$, $k_2=3.5,~p_1=-1.6$ and $p_2=1.6$ with other parameters same as in Fig. \ref{fig2ds}.}
\label{fig2ds2}
\end{figure}
The collision dynamics of dark solitons can be explored by performing an asymptotic analysis, as was done for the bright soliton collision process, which we have skipped here on considering the length of the article. From the asymptotic analysis, we find that the dark solitons appearing in the SW components undergo only elastic collision for all choices of soliton parameters, in contrast to the energy sharing collision of bright solitons in the SW components. Also, the bright solitons appearing in the LW components exhibit elastic collision as usual. But these colliding solitons experience phase-shift. By tuning the soliton parameters, one can demonstrate the collision among two dark/gray solitons or collision between a dark and a gray solitons in the SW components is elastic. Thus, irrespective of the nature of dark-soliton profile (either dark or gray) their amplitudes (intensities) remain unaltered after collision. Such an elastic collision of solitons (a dark and gray solitons in SW component and two bright solitons in the LW component) of 2-LSRI system is given in Fig. \ref{fig2ds} for $k_1=1.5,~k_2=2.5,~p_1=-1.6,~p_2=2.6$, $a_1=1$, $a_2=1.2$, $b_1=1.1$, $b_2=1.3$, $\tau_1=2$ and $\tau_2=1$. Also, in Fig. \ref{fig2ds2} we have shown the elastic collision of dark-dark ($S^{(1)}$-component), dark-gray ($S^{(2)}$-component) and bright-bright ($L$-component) solitons of 2-LSRI system.

Through dark-dark soliton collision process, one can form a bound state for the same velocity dark solitons with coinciding (different) central position(s) resulting in single (double) well type structures and they propagate like a single soliton (parallel solitons). Also, we wish to emphasis that these dark soliton bound sates do not admit periodic oscillations as in the case of bright/bright-dark solitons \cite{Kanna2012arxiv,Sakkara2013epjst}. Our procedure can be generalized to construct the dark multi-soliton solution in a straightforward manner which involve very lengthy and tedious mathematics, and the details will be presented elsewhere.

\section{Conclusions}
We have considered an integrable multicomponent long wave-short wave resonance interaction ($M$-LSRI) equation governing the dynamics of nonlinear interaction between multiple ($M$) short waves and a long wave in the context of nonlinear optics. To unravel the interesting propagation dynamics of multicomponent plane solitons we have constructed soliton solutions by using the Hirota's bilinearization method. We have briefly revisited our earlier results on the bright multi-soliton solution and demonstrated the fascinating propagation dynamics and collision processes. Particularly, we have shown that the amplitude of bright soliton appearing in the short wave components can be controlled by tuning the polarization parameters without affecting amplitude of soliton appearing in the long wave component. From the collision dynamics of solitons in $M$-LSRI system, we have identified the interesting energy sharing collision of bright solitons in the short-wave components when $M \geq 2$. The solitons in the short-wave component (for special choices of polarization parameters) can also undergo elastic collision accompanied by a phase-shift. From the dark one-soliton solution, we have observed that the nature of soliton profile (dark or gray) in the short-wave component can be controlled by tuning the soliton parameters, whereas the long-wave component supports only bright solitons. Analysis on the dark two-soliton solution reveals that the dark solitons always exhibit only elastic collisions with a phase-shift. Also, a collision between two dark/gray solitons or a collision between dark and gray solitons is also shown to be elastic. As a future study, one can construct the dark multi-soliton solution by generalizing the present algorithm and investigate the underlying dynamics.

\section*{Acknowledgments}
The work of TK is supported by Department of Science and Technology, Government of India, in the form of a major research project. KS is grateful to the support of Council of Scientific and Industrial Research, Government of India, with a Senior Research Fellowship. TK and KS also thank the principal and management of Bishop Heber College for constant support and encouragement. MV acknowledges the financial support from UGC-Dr. D. S. Kothari post-doctoral fellowship scheme. The work of ML is supported by a DST-IRPHA project. ML is also supported by a DST Ramanna Fellowship project and a DAE Raja Ramanna Fellowship.

\end{document}